\documentclass[aps,prx,reprint,superscriptaddress,nofootinbib,showkeys,floatfix]{revtex4-2}
\usepackage{graphicx}
\usepackage{xcolor}
\usepackage{amsmath,amssymb,amsfonts}
\usepackage{booktabs}
\usepackage{microtype}
\usepackage{placeins}
\usepackage[hidelinks]{hyperref}
\graphicspath{{./}}
\begin{document}

\title{Hardware-aware sampling and QPU-informed Monte Carlo with quantum annealers}

\author{Seungyeob Jwa}
\affiliation{Lattice Gauge Theory Research Center, Department of Physics and Astronomy, Seoul National University, Seoul 08826, Republic of Korea}
\author{Jangho Kim}
\email{kim.jangho1120@gmail.com}
\affiliation{Lattice Gauge Theory Research Center, Department of Physics and Astronomy, Seoul National University, Seoul 08826, Republic of Korea}
\author{Sunghee Kim}
\affiliation{Lattice Gauge Theory Research Center, Department of Physics and Astronomy, Seoul National University, Seoul 08826, Republic of Korea}
\author{Sunkyu Lee}
\affiliation{Center for Precision Neutrino Research, Department of Physics, Chonnam National University, Gwangju 61186, Republic of Korea}
\author{Weonjong Lee}
\email{wlee@snu.ac.kr}
\affiliation{Lattice Gauge Theory Research Center, Department of Physics and Astronomy, Seoul National University, Seoul 08826, Republic of Korea}

\begin{abstract}
We study how the physical representation of a logical model affects sampling by quantum annealers and its use in corrected Monte Carlo. The square-lattice Ising model provides controlled classical references. Fixed embeddings with one physical qubit per logical spin reach $36\times36$ spins. Comparisons with uniform two-qubit chains and mixed chain lengths show that matching mean energy leaves differences in magnetization and correlations. Extended-range conditional predictors improve the description of mixed embeddings, distinguishing their residual correlations from those of the tested uniform embeddings. Independent calibration checks and fixed-scale annealing scans show that temperature shifts alone do not determine distributional agreement. Normalized autoregressive models learned from QPU samples provide proposals with exact Metropolis--Hastings and multiple-try corrections on larger periodic lattices. These preserve the target distribution, although block updates alone relax critical magnetization slowly. Complementary direct-QPU updates reveal how block size and fitted-temperature correction affect acceptance and spin motion. Together, the results support prioritizing one-qubit representations and short uniform chains, while separating hardware sampling fidelity from the convergence of corrected Monte Carlo.
\end{abstract}

\maketitle

\section{Introduction}
\label{sec:introduction}

For an Ising Hamiltonian $H$, the Gibbs law is $\pi_\beta(s)=Z(\beta)^{-1}e^{-\beta H(s)}$. A quantum processing unit (QPU) based on programmable annealing provides many spin configurations, but its open-system operation does not supply a generally known Gibbs sampler~\cite{albash2018,marshall2017,vuffray2022}. Its usefulness therefore has two aspects: how faithfully the hardware samples a specified distribution, and whether its imperfect samples can improve a calculation whose target is independently controlled.

The square-lattice ferromagnet offers exact thermodynamic results, finite-volume classical references, and a stringent test of critical relaxation~\cite{onsager1944,yang1952,wolff1989,prokofev2001}. Its pairwise Hamiltonian is directly programmable without the auxiliary-variable reductions required by some other targets. The logical square lattice is nevertheless distinct from the Zephyr hardware graph~\cite{boothby2021}. This distinction makes the model useful for investigating how physical representation and calibration affect an otherwise well-defined sampling problem.

Effective-temperature inference already provides an important part of that assessment. Benedetti \emph{et al.}\ developed an instance-dependent, scale-based estimator~\cite{benedetti2016}. Raymond \emph{et al.}\ showed why local and global criteria can assign different temperatures to the same non-Gibbs output~\cite{raymond2016}. Persistent offsets can be measured and reduced~\cite{perdomo2016}, while freeze-out studies and noisy-Gibbs reconstructions identify limitations of a single-temperature description~\cite{marshall2017,vuffray2022}. Nelson \emph{et al.}\ demonstrated high-fidelity Gibbs sampling for small native instances using input-scale and duration controls~\cite{nelson2022}, and intermediate pauses provide another route to changing relaxation~\cite{marshall2019}. These results motivate testing observables beyond the statistic used to fit a temperature.

Recent work directly addresses two-dimensional thermal physics. Sathe \emph{et al.}\ studied a family including the clean square-lattice ferromagnet, using one-qubit-per-spin toroidal embeddings, disjoint parallel placements, and calibration refinement on Zephyr~\cite{sathe2026}. Teza \emph{et al.}\ used paired-qubit superspins on Pegasus for ferromagnetic tori up to 2640 logical spins, with thermometry, bias mitigation, and finite-size scaling~\cite{teza2026}. Pelofske combined chain-free parallel mappings with extensive duration and scale scans for a twelve-spin periodic axial next-nearest-neighbour Ising (ANNNI) model on Pegasus and Zephyr~\cite{pelofske2026}. Thus geometry, boundary conditions, chain overhead, active hardware, and validation criterion must all accompany a comparison of sizes or sampling quality. Direct hardware tori also differ from large periodic lattices maintained classically using block proposals.

Chain-free C1 maps use one physical qubit per logical spin. C2 uses a two-qubit chain at every site. Mixed embeddings have heterogeneous chain lengths. Ground-state-preserving embedding conditions~\cite{choi2008} do not certify finite-temperature sampling. Under ideal classical Gibbs sampling, conditioning on intact chains preserves logical relative weights, whereas generic projections of broken-chain samples can bias them~\cite{marshall2020}. Our common-scale and separately energy-calibrated C1/C2/mixed comparisons therefore test the actual hardware distributions, using energy, magnetization, and held-out conditional predictors. They distinguish temperature displacement from residual structure without assigning every discrepancy to chain breaks or a specific noise mechanism.

Validated one-to-one open blocks span $L_{\mathrm b}=3$--6, 19, and 36. The largest uses 1296 qubits and 2520 grid bonds. Zephyr's direct square-lattice construction and Glasgow subgraph methods are established~\cite{boothby2021,mccreesh2020glasgow}. We apply working-graph validation, fixed-map acquisition and repair, and parallel packing to obtain the measured realizations. Banks of 97 size-six and seven size-nineteen blocks quantify achieved capacity, while fixed-scale $20$--$500\,\mu\mathrm{s}$ scans test duration dependence and reproducibility across acquisition periods. Independent calibration checks and size-matched references separate a valid embedding from a reproducible sampling distribution.

The use of imperfect samples also builds on established neural Monte Carlo. Autoregressive models provide normalized microstate probabilities~\cite{wu2019}. Learned global or partial proposals can be corrected with Metropolis--Hastings (MH) updates~\cite{mcnaughton2020,wu2021}. Scriva \emph{et al.}\ directly demonstrated this route with annealer training data, selecting a masked autoencoder for distribution estimation (MADE) among the tested architectures~\cite{scriva2023}. We use gated recurrent unit (GRU) proposals with MH and multiple-try Metropolis (MTM) corrections to connect measured $19\times19$ blocks to periodic targets up to $512\times512$, retaining full boundary energies and distinguishing prepared-ensemble agreement from critical relaxation. A separate direct-QPU fixed-boundary (FBC) experiment compares three block sizes, two times, and two acceptance rules near criticality. Its fitted Gibbs ratio is an approximation to the hardware law, and its short trajectories diagnose acceptance and motion rather than equilibrium. Together, these tests distinguish useful proposal information from raw Gibbs fidelity, and hardware occupancy from statistically effective sampling.

The contribution of this study is a controlled connection between physical representation, residual distributional structure, and the use of measured blocks as proposals for larger lattices. We compare C1, C2, and mixed embeddings of the same logical block, test the magnetization and conditional correlations left after energy calibration, and assess learned-proposal acceptance and full-lattice relaxation in separate benchmarks. This extends the experimental comparison beyond either a fitted temperature or a valid embedding, while using established thermometry and Monte Carlo corrections~\cite{raymond2016,marshall2020,scriva2023}. The fixed-scale annealing and direct-QPU tests provide complementary checks of operating conditions and transition behaviour. Their distinct acquisitions and correction rules remain explicit.

Sections~\ref{sec:methods} and \ref{sec:reference-thermometry} establish representation and validation. Secs.~\ref{sec:hardware}--\ref{sec:obc-extension} present hardware and time controls. Sec.~\ref{sec:sampling} tests their use in periodic-lattice updates. The discussion identifies model-dependent requirements for extending the method to other Gibbs targets.

\section{Model, hardware representation, and acquisition controls}
\label{sec:methods}

\subsection{Open blocks and the periodic target}

We study the nearest-neighbour ferromagnetic Ising model on a periodic $L\times L$ lattice,
\begin{equation}
 H(s)=-\sum_{\langle ij\rangle}s_i s_j,\qquad s_i=\pm1,
 \label{eq:ising}
\end{equation}
with $J=k_{\mathrm B}=1$, $V=L^2$, and target inverse temperature $\beta=1/T$. The thermodynamic-limit critical point is $T_c=2/\ln(1+\sqrt2)=2.269185314\ldots$, or $\beta_c=0.440686794\ldots$. We reserve $L_{\mathrm b}$ and $N_{\mathrm b}=L_{\mathrm b}^2$ for an embedded block. Total magnetizations are $M=\sum_i s_i$ on the full lattice and $M_S=\sum_{i\in S}s_i$ on a block.

Open or fixed-boundary blocks avoid the extra hardware connections needed to join opposite edges. For fixed exterior spins, let $b_i=\sum_{j\notin S:\langle ij\rangle}s_j$. The conditional energy is
\begin{equation}
 H_S(s_S;b)=-\sum_{\langle ij\rangle\subset S}s_i s_j-\sum_{i\in S}b_i s_i.
 \label{eq:blockenergy}
\end{equation}
The isolated block with open boundary conditions (OBC) has $b=0$. Fixed-boundary (FBC) updates include the exterior through local fields. Replacing one block at fixed exterior changes the full periodic energy by exactly $\Delta H_S$. A Hastings update using this energy and the correct proposal ratio therefore preserves the full-lattice target even if its proposal ignores the exterior.

\subsection{Acquisition cohorts and input scale}
\label{sec:obc-scope}

The boundary-conditioned 50-context and 30-context embedding comparisons, zero-field calibration studies, fixed-scale time scans, archived zero-field GRU training data, and direct-QPU FBC trajectories are distinct cohorts. Their boundary conditions, mappings, calibration records, and acquisition periods remain attached to their results. An OBC reference is not substituted for an arbitrary FBC context. The programmed scale $\alpha$, the target $\beta$, and fitted inverse temperatures $\widehat\beta_{\mathrm X}$ are also distinct: a calibrated $\alpha_{\mathrm{cal}}$ records a selected input, not guaranteed temperature transfer to another acquisition.

\subsection{Hardware representation and fixed embeddings}
\label{sec:representation}

The annealing Hamiltonian is written schematically as
\begin{equation}
 \begin{aligned}
 \widehat{\mathcal H}(\lambda)={}&-A(\lambda)\sum_q\sigma_q^x\\
 &+B(\lambda)\left[\sum_q h_q\sigma_q^z+
 \sum_{(q,q')\in C}J_{qq'}\sigma_q^z\sigma_{q'}^z\right],
 \end{aligned}
 \label{eq:annealing-hamiltonian}
\end{equation}
Here $q$ labels a physical qubit, $\sigma_q^{x,z}$ are Pauli operators, and $C$ is the set of active hardware couplers. The coefficients $h_q$ and $J_{qq'}$ are programmed physical biases and couplings, while $A(\lambda)$ and $B(\lambda)$ set the transverse and Ising energy scales. The reduced annealing time is $\lambda=t/t_a$, with duration $t_a$. The logical scale $\alpha$ multiplies both intended couplings and boundary fields. Neither it nor the annealing duration specifies a known output temperature~\cite{benedetti2016,marshall2017}. We use the recorded active Zephyr graph, whose nominal degree is 20, to realize the degree-four square grid~\cite{boothby2021}.

For the physical chain $C_i$ representing logical site $i$, define
\begin{equation}
 \ell_i=|C_i|,\qquad \ell_{\max}=\max_i\ell_i.
 \label{eq:chainlength}
\end{equation}
Chain-free (C1) maps have $\ell_i=1$. Uniform-length-two (C2) maps have $\ell_i=2$ at every site. Mixed-chain maps have heterogeneous lengths. Here $\ell=2$ means a coupled pair representing each logical spin, not a geometric spacing of two. C1 removes internal chain breaks, while C2 controls chain-length heterogeneity. Neither construction by itself establishes sampling accuracy.

\subsection{Search, validation, and parallel packing}
\label{sec:chainfree-search}

For a logical grid $G_{L_{\mathrm b}}=(V_{L_{\mathrm b}},E_{L_{\mathrm b}})$ and working graph $G_{\mathrm{hw}}=(Q,C)$, a C1 mapping obeys
\begin{align}
 \varphi &:V_{L_{\mathrm b}}\longrightarrow Q,
 &\varphi(i)&\ne\varphi(j)\quad(i\ne j),\nonumber\\
 (i,j)&\in E_{L_{\mathrm b}}
 &\Longrightarrow\quad &(\varphi(i),\varphi(j))\in C.
 \label{eq:chainfree-subgraph}
\end{align}
This non-induced subgraph uses $L_{\mathrm b}^2$ qubits and $2L_{\mathrm b}(L_{\mathrm b}-1)$ required couplers. Unused hardware edges are programmed to zero. Missing logical bonds are not replaced by chains.

Small-grid searches use seeded, constraint-preserving backtracking with adjacency and forward checks. Larger grids use Glasgow subgraph search through \texttt{minorminer.subgraph.find\_subgraph}~\cite{mccreesh2020glasgow}, with portfolios of variable orderings, connected windows, and complete-grid or anchor-completion attempts. Central windows favour compact placements. Regional searches supply alternatives. Crops of verified larger grids retain their parent and origin because they can overlap. Every accepted map is checked for complete logical coverage, one physical qubit per logical spin, active qubits, and all required bonds, then frozen with graph and mapping hashes. A graph change triggers revalidation. Search timeout does not prove nonexistence, and the largest returned map is not a maximum-size certificate. Implementation details are in Appendix~\ref{app:chainfree-search}.

The direct-QPU campaign additionally maximized the number of disjoint fixed blocks per parallel bank. Seeded CPU workers shared feasible layouts, generated candidates, and improved packings by joint qubit assignment and block removal/refill. Selection prioritized bank width before compact-centre rank. On graph \texttt{01b03fc8dc} (4560 active qubits, 41205 couplers), the best verified banks contain 97 size-six and seven size-nineteen blocks. A separately repaired size-thirty-six map provides a one-block bank. The widths are achieved search results, not proven global or candidate-pool optima. Complete-bank occupancies are 76.58\%, 55.42\%, and 28.42\%. Actual mean request occupancies are 63.68\%, 48.49\%, and 28.42\%, because not every request fills its bank. Appendix~\ref{app:new-resource-accounting} gives the acquisition counts and cost definitions. Search and packing are classical computations and consume no QPU samples.

Available hardware differs substantially across the closest experiments. Sathe \emph{et al.}\ used Zephyr \texttt{Advantage2\_prototype2.6}~\cite{sathe2026}. Their size-six toroidal C1 bank contains 16 blocks, whereas our size-six open bank contains 97. Chip size and boundary connectivity both affect this comparison. Teza \emph{et al.}'s largest torus contains 2640 logical spins represented by 5280 physical qubits on Pegasus~\cite{teza2026}. Pelofske reports 337 Pegasus and 204 Zephyr placements of a twelve-spin periodic ANNNI model, while explicitly identifying the tabulated Zephyr qubit count as newer than the acquisition graph~\cite{pelofske2026}. These quantities describe different physical resources and logical tasks. They do not establish an embedding-algorithm ranking or independent samples per unit cost.

\subsection{Geometry and acquisition controls}
\label{sec:embedding-geometry}

\begin{figure*}[t]
\centering
\includegraphics[width=0.96\textwidth]{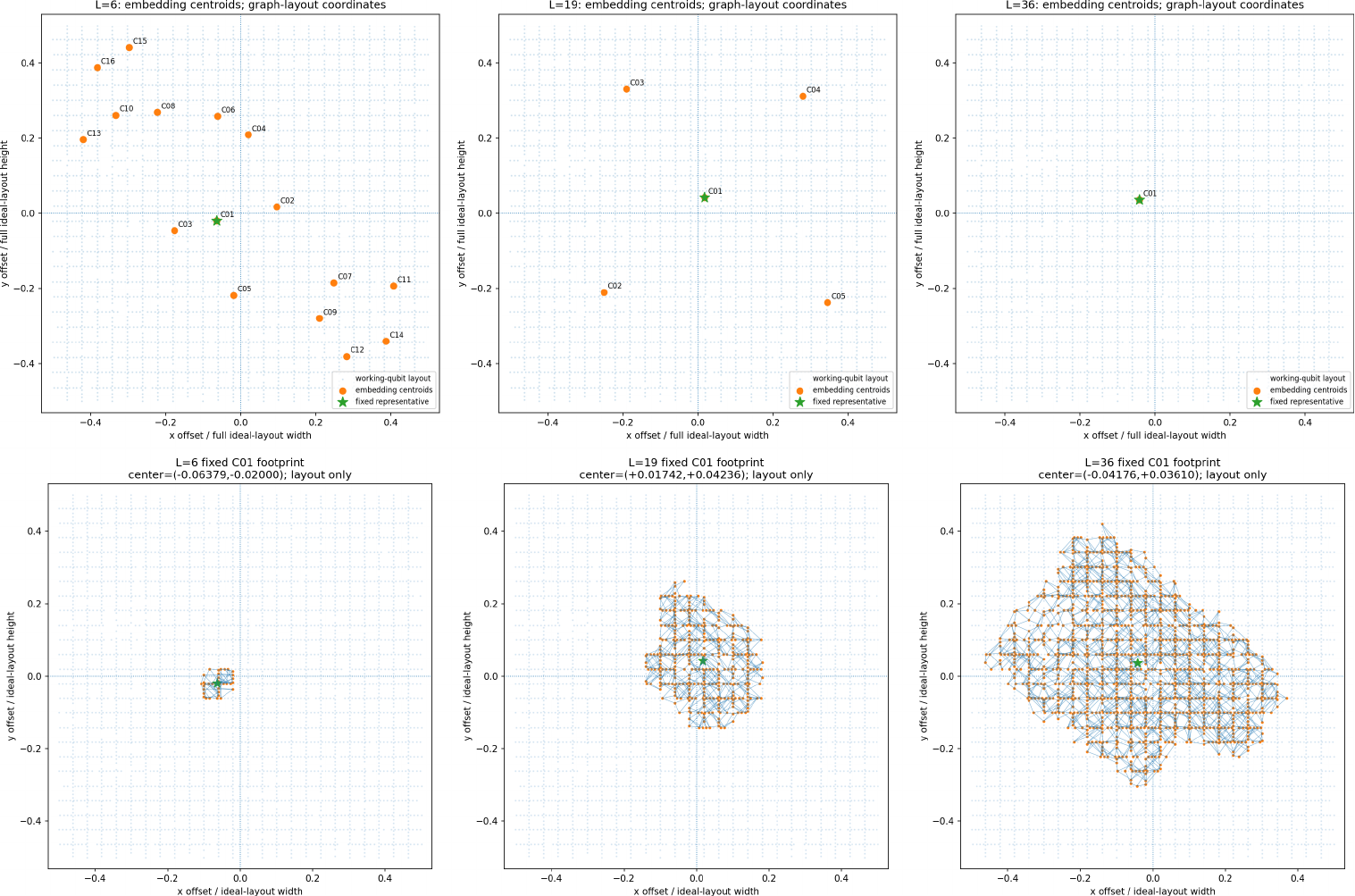}
\caption{Physical layouts of C1 blocks of side length 6, 19, and 36, from left to right. The upper row locates embedding centres and the fixed representative on the active graph. The lower row shows the qubits and couplers of each fixed representative. All panels use normalized graph-layout coordinates, not physical distances. The panels do not establish a geometry--accuracy correlation.}
\label{fig:obc-geometry}
\end{figure*}

For normalized chip-drawing coordinates, we record
\begin{equation}
 (\Delta x,\Delta y)=(\overline x-\tfrac12,\tfrac12-\overline y),
 \quad d_{\mathrm{centre}}=\sqrt{\Delta x^2+\Delta y^2}.
 \label{eq:chip-centre}
\end{equation}
These are layout coordinates, not fabrication distances. Radius of gyration measures spread independently of centre distance. Figure~\ref{fig:obc-geometry} shows representative C1 layouts requiring 36, 361, and 1296 qubits and 60, 684, and 2520 grid bonds. Only one size-thirty-six placement is shown. Geometry is a selection descriptor, not a demonstrated predictor of sampling quality.

\subsection{Gauge and placement controls}
\label{sec:placement-controls}
A quantum-machine instruction (QMI) is one programmed acquisition returning multiple reads. Gauges $g_i=\pm1$ encode $z_{\varphi(i)}=g_i s_i$ with $J_{\varphi(i)\varphi(j)}=-\alpha g_ig_j$. For a C1 block with boundary input, $h_{\varphi(i)}=-\alpha g_i b_i$. Zero-field acquisitions have $h_q=0$. We undo the gauge before analysis and disable automatic coefficient rescaling. Typical calibration conditions use four gauges and two repeats, initially 250 reads/QMI. Independent checks and later time scans use 500. Acquisition periods remain distinct from QMI repeats and neural training epochs.

Calibration selects a scale. A fresh check acquires new reads at that fixed scale. It is independent of scale selection, not necessarily of the hardware state. Sequential acquisition uses one map, whereas disconnected parallel acquisition uses disjoint blocks with zero programmed inter-block coupling. Such blocks can remain statistically correlated, so their physical-QMI identity is preserved in uncertainty estimates. Common-scale placement comparisons match size, boundary condition, scale, time, and gauge design. Energy-calibrated comparisons instead select $\alpha$ separately for each placement. Signed-magnetization and local-response diagnostics are retained in Appendix~\ref{app:obc-gauge}. Residual asymmetry is not by itself evidence of a defective qubit. These controls separate temperature shifts from the remaining observable discrepancies defined below.

\section{Classical references and statistical assessment}
\label{sec:reference-thermometry}

\subsection{Finite-block reference hierarchy}
\label{sec:classical-hierarchy}

A reference must match the logical Hamiltonian, size, and boundary condition. Infinite-volume Ising results therefore do not replace finite-OBC energy or magnetization distributions. A finite open block has no thermodynamic singularity at $T_c$. Its response peaks depend on size and observable. For $L_{\mathrm b}=3$--6, integer row-wise dynamic programming gives the joint density of states (DOS) $g(E,M_S)$: $E=-2L_{\mathrm b}(L_{\mathrm b}-1)+2n_{\mathrm{dw}}$ and $M_S=2n_\uparrow-N_{\mathrm b}$. Here $n_{\mathrm{dw}}$ counts disagreeing nearest-neighbour bonds and $n_\uparrow$ counts up spins. This exact state counting is equivalent to enumeration without traversing all $2^{36}$ configurations at size six.

For size nineteen, factorized row transfer uses internal row energy $u(r)=-\sum_jr_jr_{j+1}$ and inter-row energy $v(r,r')=-\sum_jr_jr'_j$:
\begin{align}
 F_1(r)&=e^{-\beta u(r)},\nonumber\\
 F_{k+1}(r')&=e^{-\beta u(r')}\sum_rF_k(r)e^{-\beta v(r,r')},\nonumber\\
 Z_{L_{\mathrm b}}(\beta)&=\sum_rF_{L_{\mathrm b}}(r).
 \label{eq:row-transfer-reference}
\end{align}
It provides $Z$, mean energy, and heat capacity at computed temperature nodes up to numerical arithmetic. Interpolation is separate, and these exact energy quantities do not provide an exact joint energy--magnetization law. The size-thirty-six pipeline uses Monte Carlo (MC) and the weighted histogram analysis method (WHAM), with independent construction and validation trajectories. Table~\ref{tab:reference-hierarchy} specifies what is exact in the implemented analysis. It does not assert that exact finite-rectangle algorithms are unavailable at larger sizes.

\begin{table*}[t]
\centering
\caption{References for zero-field open blocks. Exact thermodynamic quantities do not make an empirical or reconstructed distribution exact. All comparisons retain the reference revision used.}
\label{tab:reference-hierarchy}
\small
\begin{tabular}{p{0.07\textwidth}p{0.31\textwidth}p{0.47\textwidth}}
\toprule
$L_{\mathrm b}$ & Exact calculation & Remaining qualifications\\
\midrule
3--6 & Integer joint DOS $g(E,M_S)$ and its moments & QPU comparisons retain finite-sample uncertainty\\
19 & Row-transfer $Z$, $\langle E\rangle$, $C_V$ at computed nodes & Interpolation, MC magnetization moments, and empirical/WHAM energy laws retain their own uncertainty\\
36 & No exact calculation used in this pipeline & MC/WHAM require mixing, histogram overlap, energy support, and numerical-convergence checks\\
\bottomrule
\end{tabular}
\end{table*}

Reference simulations have a specified, known $\beta$. Their statistical uncertainty is distinct from the effective temperature inferred for QPU samples. Support, overlap, and mixing must be checked separately from numerical WHAM convergence. Observations outside the reference support are retained rather than truncated. Construction and validation details are in Appendix~\ref{app:obc-reference}.

\subsection{Local and global temperature estimates}
\label{sec:obc-thermometry}

For the logical local field $\eta_i=b_i+\sum_{j\in S:\langle ij\rangle}s_j$, a homogeneous Gibbs model has
\begin{equation}
 \Pr(s_i\mid \eta_i,\beta)=[1+e^{-2\beta s_i\eta_i}]^{-1}.
 \label{eq:conditional}
\end{equation}
Using all logical sites and reads, the scalar pseudolikelihood (PL) fit is
\begin{equation}
 \widehat\beta_{\mathrm{PL}}=\mathop{\mathrm{arg\,min}}_{\beta'}
 \sum_{c,r,i}\log[1+e^{-2\beta's_i^{(c,r)}\eta_i^{(c,r)}}],
 \label{eq:pseudotemperature-fit}
\end{equation}
where $c$ labels boundary contexts, $r$ labels reads, and $i$ labels logical sites. The local field $\eta_i$ is calculated from the neighbouring spins and boundary input. It is distinct from the programmed physical bias $h_q$ and is not multiplied by $\alpha$ again. For OBC, $b=0$. Fitted temperatures are operational logical quantities, not measurements in kelvin~\cite{benedetti2016,raymond2016}.

For a zero-field DOS, the equilibrium energy law is
\begin{equation}
 p_\beta(E)=\frac{g_{L_{\mathrm b}}(E)e^{-\beta E}}{Z_{L_{\mathrm b}}(\beta)},
 \quad Z_{L_{\mathrm b}}(\beta)=\sum_Eg_{L_{\mathrm b}}(E)e^{-\beta E}.
 \label{eq:obc-energy-law}
\end{equation}
The primary zero-field calibration coordinate is
\begin{equation}
 \widehat\beta_{\mathrm{JS}}=\mathop{\mathrm{arg\,min}}_{\beta'\in\mathcal B}
 D_{\mathrm{JS}}[\widehat P(E),p_{\beta'}(E)],
 \label{eq:obc-js-temperature}
\end{equation}
with recorded fit interval $\mathcal B$. We use the Jensen--Shannon (JS) divergence with natural logarithms, $D_{\mathrm{JS}}(P,Q)=\tfrac12D_{\mathrm{KL}}(P\Vert R)+\tfrac12D_{\mathrm{KL}}(Q\Vert R)$ with $R=(P+Q)/2$, where $D_{\mathrm{KL}}$ is the Kullback--Leibler (KL) divergence~\cite{lin1991}. Its finite value under support mismatch does not validate an incomplete reference. A reported fit requires adequate support and an identifiable minimum within the recorded bounds, rather than merely a finite optimizer return. The original boundary-conditioned comparison instead applies JS to the discrete joint $(E,M_S)$ distributions, both pooled and averaged over shared boundaries.

The same-DOS energy estimate solves $\langle E\rangle_{p_{\widehat\beta_E}}=\langle E\rangle_{\mathrm{data}}$. With identical support, weights, and fit bounds, forward-KL minimization gives the same equation and is an implementation cross-check, not an independent thermometer. The separately tabulated energy estimate $\widehat\beta_{E,\mathrm{table}}$ retains its interpolation and reference provenance. Estimator disagreement can reveal a noncanonical distribution but does not identify its physical cause. Fitting a reference temperature also differs from changing $\alpha$ and acquiring new samples.

\subsection{Observable residuals and uncertainty}
\label{sec:thermometry-criteria}
\label{sec:obc-shape}

For any observable $O$, we separate target mismatch from a temperature-adjusted residual:
\begin{align}
 \widehat O-O_{\mathrm{ref}}(\beta)
 &=\big[\widehat O-O_{\mathrm{ref}}(\widehat\beta_{\mathrm{JS}})\big]\nonumber\\
 &\quad+\big[O_{\mathrm{ref}}(\widehat\beta_{\mathrm{JS}})-O_{\mathrm{ref}}(\beta)\big].
 \label{eq:shape-decomposition}
\end{align}
A held-out version estimates $\widehat\beta$ on one repeat set and evaluates observables on the other, then exchanges roles. The target discrepancy and estimator difference are
\begin{equation}
 \delta_{\beta,\mathrm X}=\frac{\widehat\beta_{\mathrm X}-\beta}{\beta},
 \quad\Delta\beta_{\mathrm X-\mathrm{JS}}=\widehat\beta_{\mathrm X}-\widehat\beta_{\mathrm{JS}}.
 \label{eq:beta-comparison-notation}
\end{equation}
For block observables, we use $e_{\mathrm b}=E/N_{\mathrm b}$, $m_{\mathrm b}=M_S/N_{\mathrm b}$, $\langle|m_{\mathrm b}|\rangle$, and the pooled-moment Binder cumulant
\begin{equation}
 U_4=1-\frac{\langle m_{\mathrm b}^4\rangle}{3\langle m_{\mathrm b}^2\rangle^2}.
 \label{eq:obc-binder}
\end{equation}
Here $E=H_S(s_S;b)$ is the total block energy, with $b=0$ for OBC. For full-lattice observables, we use $e=H(s)/V$ and $m=M/V$, and apply the same Binder definition to $m$. The block magnetization labelled $M_B$ or $M$ in the embedding figures is $M_S$ in the text. Mean-energy matching leaves magnetization, fluctuations, and spatial structure unconstrained. A small empirical JS value requires a finite-sample baseline before it can be interpreted as unusually good agreement. Similarly, $P(E,M_S)$ does not determine probabilities within a macrostate shell. Agreement of projected observables or inclusion of a target in a confidence interval is not full-distribution equivalence.

Intervals are 95\% confidence intervals unless marked as standard errors. The original 50-boundary comparison resamples complete contexts: 2000 refits for PL and 200 paired resamples for mean boundary-wise JS. The fixed-gauge OBC studies resample whole QMIs within each gauge, keeping parallel blocks from the same physical QMI together. Temperature estimators compared on the same data use shared bootstrap draws, so their differences retain covariance. Conditional-model details are in Appendix~\ref{app:conditional}.

The reported intervals condition on the supplied classical reference. Reference uncertainty and later-period variation are not automatically included. Two repeats per gauge also limit held-out precision, so available held-out residuals are reported as point estimates without a well-resolved interval. These uncertainty conventions distinguish reproducible analysis, calibration transfer, and genuine agreement with the specified Gibbs law.

\section{Effects of embedding class on the sampled distribution}
\label{sec:hardware}

\subsection{Direct QPU comparison under the same sampling conditions}

The boundary-conditioned comparison uses an open $19\times19$ square graph on an Advantage2 working graph. The C1 map occupies 361 physical qubits. The mixed-chain map assigns one, two, three, and four physical qubits to 226, 96, 31, and 8 logical spins, respectively, giving $\ell_{\max}=4$ under Eq.~\eqref{eq:chainlength}. This distribution describes the heterogeneous representation more fully than its maximum alone.

We selected random $19\times19$ windows from equilibrated periodic $128\times128$ configurations at $T_c$ and recorded the 76 exterior spins defining the boundary fields $b_i$. Programming these fields as local biases on boundary spins implements the conditional Hamiltonian $H_S(s_S;b)$ of Eq.~\eqref{eq:blockenergy} on an open embedded graph. The fields act at the block boundary. The logical interior retains the same nearest-neighbour square lattice in both physical representations. Fifty shared boundary contexts and 1,000 raw reads per context give 50,000 reads per embedding, or 100,000 in total. The common scale $\alpha=0.0622$ was selected by interpolating preliminary mixed-embedding inverse-temperature estimates to $\beta_c$. These data also support the proposal tests near $T_c$.

\begin{figure*}[!t]
\centering
\includegraphics[width=0.98\textwidth]{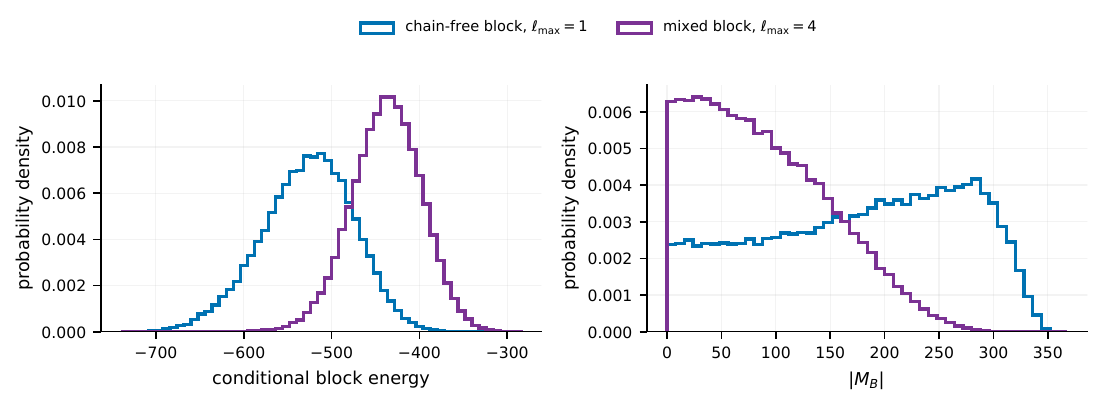}
\caption{Energy and absolute magnetization distributions at the common scale $\alpha=0.0622$, using 50,000 chain-free and 49,604 unbroken mixed reads from 50 shared boundaries. The pooled JS divergence is 0.435. Its boundary-wise mean is 0.652 (95\% interval, 0.647--0.656). Effective temperatures differ. Fig.~\ref{fig:calibration-histograms} shows the separate energy-calibrated comparison.}
\label{fig:postselection}
\end{figure*}

\begin{table*}[!t]
\centering
\caption{Conditional inverse temperatures from all logical spins and unbroken reads. The 95\% intervals resample whole boundary contexts and refit $\widehat\beta_{\mathrm{PL}}$.}
\label{tab:hardware}
\small
\begin{tabular}{lrr}
\toprule
Embedding & $\widehat\beta_{\mathrm{PL}}$ & 95\% interval\\
\midrule
Chain-free & 0.45580 & [0.45359,0.45828]\\
Mixed, $\ell_{\max}=4$ & 0.43643 & [0.43474,0.43810]\\
\bottomrule
\end{tabular}
\end{table*}

At this common programmed scale, C1 has lower energy and larger absolute magnetization than the mixed embedding (Fig.~\ref{fig:postselection}). The joint distributions remain separated within individual shared boundaries. All logical spins and bonds, including those joining different chain lengths, enter the comparison. Reads with any final chain break are excluded from the direct distribution and conditional-response analyses. This changes the mean observables only slightly. The different conditional temperatures in Table~\ref{tab:hardware} are consistent with the observed ordering. Thus the common-scale comparison does not by itself establish a discrepancy beyond a temperature shift. The next experiment calibrates mean energy separately before testing residual distributional differences.

\subsection{Comparison of multiple embeddings after energy calibration}

A separate acquisition tests recurrence across physical placements. Each chain-free embedding uses a distinct set of 361 qubits. Mixed embeddings have mean chain lengths 1.100--1.850 and maxima 3--11. For each embedding, seven scales from $\alpha=0.05$ to 0.14 were measured using eight shared contexts and 300 reads per point. Interpolation between neighbouring points selected the scale targeting a common classical conditional mean energy. This adjusts one global $\alpha$ for both logical couplings and boundary fields. New independent-check acquisitions then used 30 contexts and 1,000 reads per context, excluding broken reads from the primary analysis. Calibration selects a programmed scale. The independent check tests observables in new data. A matched energy mean does not establish a common thermodynamic temperature when the acquired distributions are not Boltzmann.

\begin{figure*}[!t]
\centering
\includegraphics[width=0.98\textwidth]{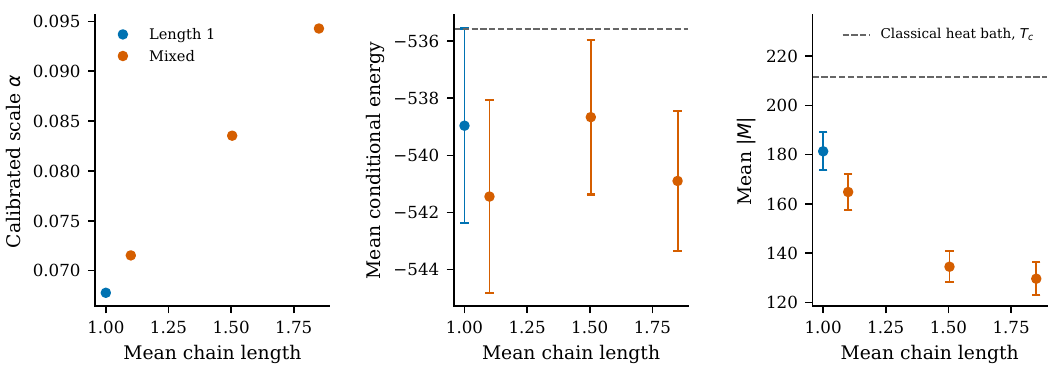}
\caption{Selected $\alpha$, mean conditional energy, and mean absolute magnetization. Blue averages chain-free placements. Orange retains three representative mixed placements, one per pair of similar mean chain lengths. Dashed lines mark the energy target ($-535.58$) and classical magnetization reference (211.52). Error bars are standard errors over boundary contexts.}
\label{fig:embedding-scaling}
\end{figure*}

Figure~\ref{fig:embedding-scaling} summarizes representative placements. Longer mean mixed-chain lengths generally require larger $\alpha$ to approach the energy target. At the selected scales, mean energies lie near that target with residual offsets, while absolute magnetization decreases with mean chain length. Energy measures nearest-neighbour alignment, including boundary bonds. Magnetization measures alignment across the block. Their different responses therefore test structure that matching a single energy mean does not constrain.

\begin{figure*}[!t]
\centering
\includegraphics[width=0.98\textwidth]{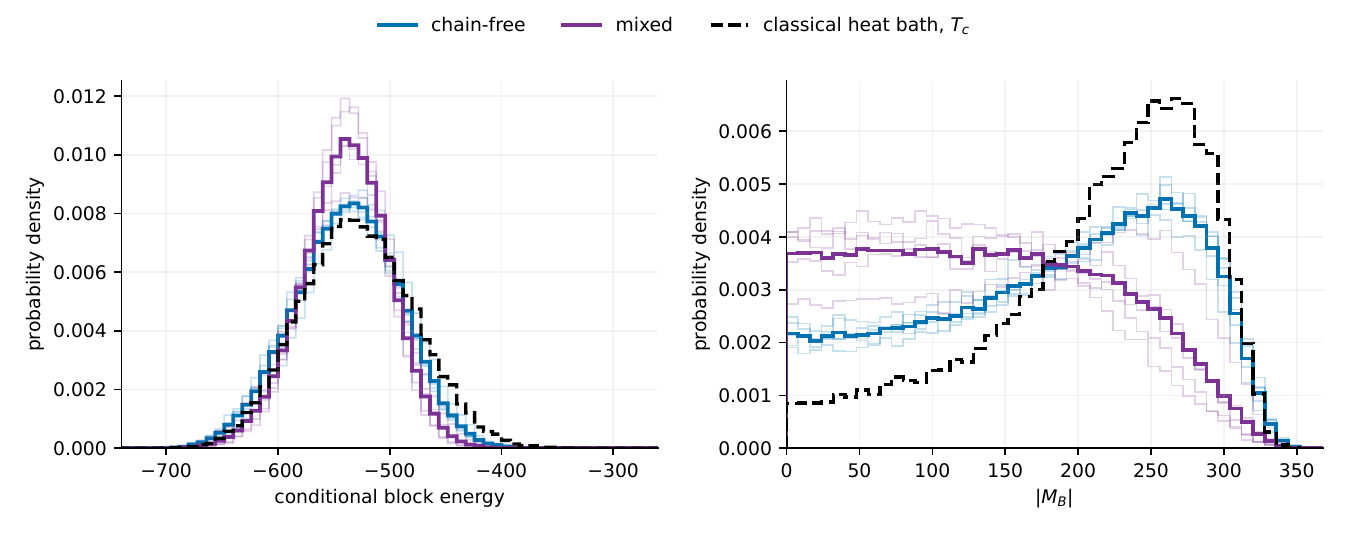}
\caption{Distributions after separate energy calibration, compared with the classical heat bath reference at $T_c$ on the same 30 boundaries (black dashed line). Thin curves show individual embeddings. Thick curves are equally weighted means of their normalized histograms. QPU reads with broken chains are excluded.}
\label{fig:calibration-histograms}
\end{figure*}

Figure~\ref{fig:calibration-histograms} compares the ten individually calibrated embeddings with a classical heat bath reference at $T_c$ on the same 30 boundaries. The reference uses 1,000 replicas per boundary, half initially ordered and half random, each evolved for 2,500 checkerboard heat bath sweeps before one configuration is recorded. Although the two QPU groups have close mean energies, mixed embeddings retain a narrower energy distribution. C1 is closer to the reference in mean absolute magnetization, but both groups underestimate it. Matching mean energy therefore does not align energy fluctuations or magnetization distributions.

The calibration target and independent-check reference average over different boundary sets. Both QPU groups have slightly lower mean energies than the latter reference, so residual energy offsets remain relevant when interpreting these distributional differences. In a separate check, one temperature per embedding was fitted to its mean conditional energy on 24 boundaries, retaining the programmed boundary fields, and the resulting Gibbs model was tested on the other six. Mean-energy offsets were small but nonzero. The fitted models still had higher absolute magnetization than the QPU samples, with a larger discrepancy for mixed embeddings.

Additional chain-strength and annealing-time tests show that fewer final chain breaks need not give closer magnetization agreement (Appendix~\ref{app:intervention}).

\subsection{Effects of chain length in uniform embeddings}
\label{sec:uniform-control}

The uniform-length-two control separates uniform chain length from the heterogeneous representations above. Every logical spin in its $19\times19$ block uses a two-qubit chain. Multiple uniform and mixed placements on the same Zephyr processor were calibrated to the mean energy of a C1 QPU reference, then measured anew. The reference scale was 0.0622. Selected scales were approximately 0.0955--0.0970 for uniform-length-two and 0.0714--0.0855 for mixed embeddings. Measurements used $t_a=20\,\mu\mathrm{s}$, chain strength 1, disabled automatic scaling, and 1,000 reads for each of 30 boundary contexts. Separate annealing-time tests did not substantially remove the discrepancy from the classical magnetization reference (Appendix~\ref{app:intervention}).

Here the reference is a C1 QPU sample, not a classical equilibrium calculation. Uniform and final mixed measurements were acquired separately, the latter after an additional calibration adjustment, and each is compared with its own C1 reference. We calculate
\begin{align}
\Delta E &= \langle E\rangle_{\mathrm{test}}-\langle E\rangle_{\ell=1}, \nonumber\\
\Delta|M_S| &= \langle |M_S|\rangle_{\mathrm{test}}-\langle |M_S|\rangle_{\ell=1},
\end{align}
where \textit{test} denotes uniform-length-two or mixed embedding. Averages weight the 30 boundaries equally and use total block observables, without division by 361. Negative $\Delta|M_S|$ denotes a magnetization deficit relative to the corresponding reference.

Mixed embeddings have the larger magnetization deficit, with the contrast interval excluding zero, whereas the energy contrast remains unresolved. This compares deviations from the respective references, not unpaired means from different acquisition periods. Intervals resample whole boundaries while keeping test and reference measurements paired. They condition on the selected $\alpha$ and exclude uncertainty from repeating the calibration. Uniform-length-two is therefore closer to its C1 reference in magnetization for the tested placements. Differing placements and acquisition periods prevent attributing that improvement to chain uniformity alone.

\begin{figure*}[t]
\centering
\includegraphics[width=0.96\textwidth]{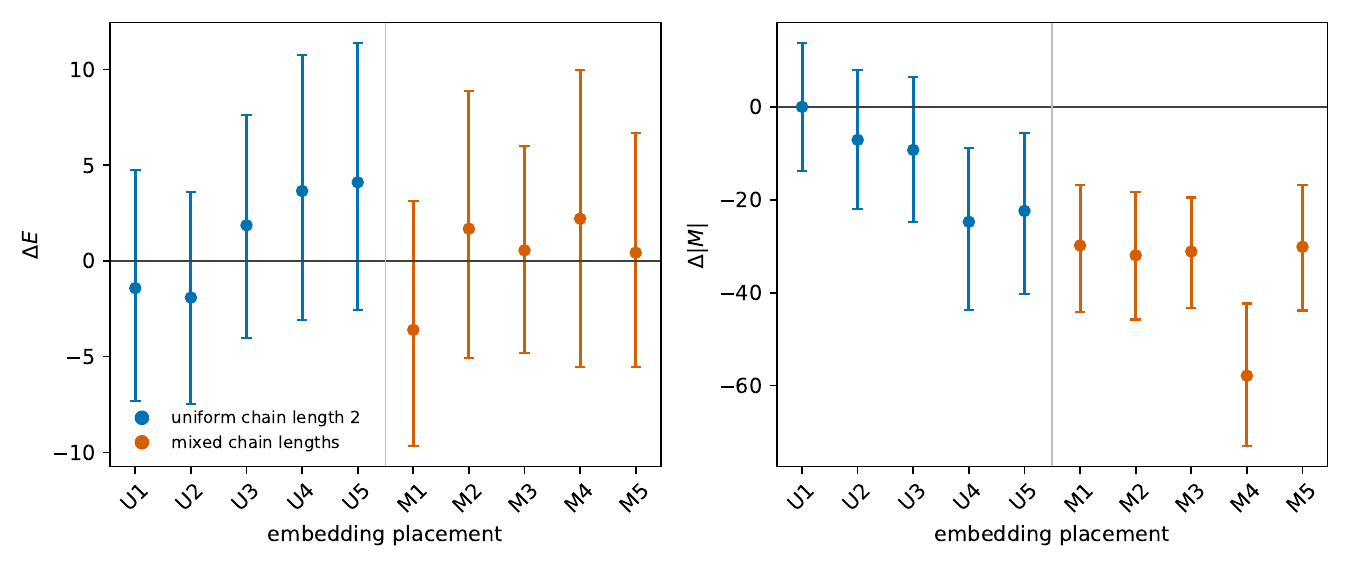}
\caption{Total-energy and absolute-magnetization differences after mean-energy calibration. U1--U5 are uniform-length-two and M1--M5 mixed embeddings, each compared with its C1 QPU reference. Error bars are 95\% paired-boundary resampling intervals. Zero denotes agreement with the reference mean.}
\label{fig:uniform-control}
\end{figure*}

The separate single-placement control is summarized in Appendix~\ref{app:conditional}. The learned-proposal tests remain distinct: Fig.~\ref{fig:acceptance} and Sec.~\ref{sec:temperature-acceptance} evaluate proposals at their own fitted temperatures, whereas Table~\ref{tab:uniform-temperature} compares them at one common target temperature.

\subsection{\textcolor{black}{Interaction range in uniform and mixed embeddings}}
\label{sec:chain-range}

\textcolor{black}{We next ask whether the measured distributions can be described by allowing the nearest neighbour couplings and site fields to vary across the block, or whether predictors beyond nearest neighbours improve that description. We use the Zephyr acquisition containing a length one reference placement and multiple uniform length two and mixed placements on the same 30 boundaries. This is the initial calibrated panel, before the additional mixed calibration used for Fig.~\ref{fig:uniform-control}. The logical scales were selected separately for each placement. The new analysis does not assume equal fitted temperatures or exact mean energy matching. Only reads without broken chains are retained.}

\textcolor{black}{For each placement, we fit symmetric, dimensionless couplings $K_{ij}$, site fields $u_i$, and responses to the fixed exterior spins. The nearest neighbour model allows every programmed bond and site to have a different coefficient. We then include all internal pairs within Manhattan distance two and, in a third model, distance three. Distance two includes diagonal neighbours and pairs two lattice spacings apart along an axis. Each model is fitted on four of five groups of complete boundary contexts and tested on the remaining group, using the same groups for all placements. Thus a distant spin must improve prediction beyond the locally varying nearest neighbour model on boundaries excluded from fitting. Appendix~\ref{app:range-fit} gives the fitting details.}

\textcolor{black}{We evaluate prediction using the conditional negative log likelihood (NLL),}
{\color{black}
\begin{equation}
\begin{aligned}
\mathrm{NLL}_{\mathrm{cond}}
&=-\frac{1}{N_{\mathrm{ctx}}}\sum_{c=1}^{N_{\mathrm{ctx}}}\frac{1}{n_cN_{\mathrm b}}
  \sum_{r=1}^{n_c}\sum_{i\in S}\\[-2pt]
&\qquad \log p_{\vartheta_c}\!\left(
s_i^{(c,r)}\mid s_{S\setminus i}^{(c,r)},\boldsymbol\sigma_c
\right).
\end{aligned}
\label{eq:conditional-nll}
\end{equation}
where $N_{\mathrm{ctx}}$ is the number of evaluation boundary contexts, $c$ indexes these contexts, $n_c$ is the number of evaluation samples in context $c$, and $r$ indexes the samples. The set $S$ contains the logical block sites, $N_{\mathrm b}=|S|$ is its number of spins, and $i$ indexes a site. The observed spin is $s_i^{(c,r)}$, while $s_{S\setminus i}^{(c,r)}$ denotes all remaining block spins and $\boldsymbol\sigma_c$ is the vector of exterior spins for context $c$. In this conditional fit, the exterior spins $\sigma_a$ are retained individually for their fitted boundary-bond responses, as specified in Appendix~\ref{app:range-fit}, rather than reduced to a single common response to their sum. The conditional probability $p_{\vartheta_c}$ is assigned to the observed spin given the remaining spins and boundary. Its coupling and field parameters $\vartheta_c$ are fitted with the evaluation fold containing context $c$ excluded. Contexts in the same fold share the fitted model. The logarithm is natural. Averaging the negative log probabilities over spins and samples with equal weight for each context gives smaller NLL for better prediction. We define predictive gain as
\begin{equation}
\Delta\mathrm{NLL}=\mathrm{NLL}_{\mathrm{shorter}}-\mathrm{NLL}_{\mathrm{extended}}.
\label{eq:nll-gain}
\end{equation}
where the subscripts $\mathrm{shorter}$ and $\mathrm{extended}$ denote the shorter-range and extended models being compared. Both are evaluated on the same samples, so $\Delta\mathrm{NLL}>0$ means that the additional terms improve prediction on boundaries not used for fitting.
}

\begin{figure}[!htbp]
\centering
\includegraphics[width=\linewidth]{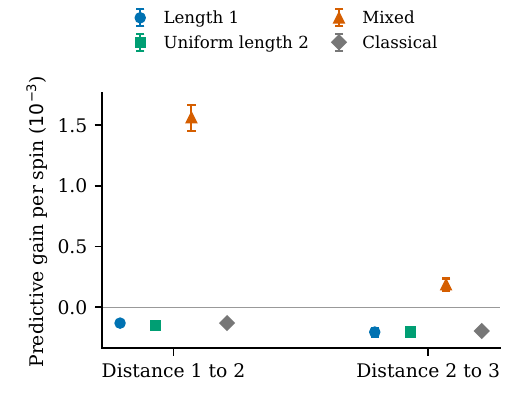}
\caption{\textcolor{black}{Improvement in conditional NLL per spin on held-out boundaries as the interaction range increases. Positive values favour the extended model. Bars show 95\% context-resampling intervals. Each embedding class is represented by its lowest stored placement index, alongside a classical nearest neighbour control.}}
\label{fig:chain-range}
\end{figure}

\textcolor{black}{Figure~\ref{fig:chain-range} shows one placement of each type, and the full comparison gives the same qualitative distinction. Extending the nearest neighbour model to distance two improves prediction for every mixed placement, whereas neither length one nor any uniform length two placement has a positive 95\% interval excluding zero. Extending further to distance three again improves every mixed placement, while gains are negative for length one and all uniform length two placements. All fits satisfy the numerical convergence checks.}

\textcolor{black}{The classical control uses the same boundaries and a nearest neighbour heat bath ensemble at $T_c$. Its gains are negative for both extensions. It tests the fitting procedure on a known nearest neighbour model, rather than matching the temperature of each QPU distribution. Negative gain means that the added terms increase NLL on the evaluation data and worsen prediction relative to the simpler model, consistent with overfitting to a finite sample.}

\textcolor{black}{The predictive comparison supports a distinction between uniform and mixed embeddings, rather than a distinction between having and not having chains. Additional pair terms improve the description of the tested mixed samples even after fitting spatially varying nearest neighbour couplings and fields. No corresponding benefit is resolved for uniform length two. These are effective dependencies in the logical samples, not additional couplers programmed on the QPU. The result motivates short, uniform chains as an alternative when one physical qubit per logical spin is unavailable.}

\section{Calibration transfer and annealing-time dependence}
\label{sec:obc-extension}

\subsection{Independent calibration checks}
\label{sec:obc-transfer}

Interpolation within a measured response curve and transfer to a new acquisition are different tests. We select a candidate $\alpha$ only from a finite, bracketed response and evaluate it using new reads that did not select the scale. A failed check remains part of the comparison. An extrapolated probe is not a calibrated value. Holding the logical map fixed removes embedding reselection, but does not hold the hardware response fixed.

For two $L_{\mathrm b}=19$ placements at $T=3$ and $20\,\mu\mathrm{s}$, interpolation bracketed the target $\beta=1/3$, yet the independent JS estimates were 0.280666 and 0.285410, approximately 15.8\% and 14.4\% below target. Reanalysis reproduced these values, separating numerical reproducibility from calibration transfer. The corresponding $L_{\mathrm b}=36$ checks gave deviations of $+2.05\%$ at $T=2.17$ and $-0.066\%$ at $T=2.27$. The latter target falls inside its reported interval, but this is not an equivalence test. These changes cannot be assigned uniquely to hardware drift, placement, or reference error.

\subsection{Time dependence at calibrated and fixed input scales}
\label{sec:obc-anneal}
\label{new-anneal:fixed-scale}

The initial time study independently checked a separately selected $\alpha(T,t_a)$ at each duration. In its $20$--$100\,\mu\mathrm{s}$ comparisons, same-DOS mean-energy estimates were close to JS estimates, while scalar pseudolikelihood point estimates were higher. These checks probe calibration and transfer together: changes of scale and acquisition period prevent their interpretation as a time-only intervention. Nor did they identify one optimum common to all block sizes.

We therefore performed a separate fixed-scale study. For $L_{\mathrm b}=6$ and 19, SHORT used $20,40,60,80,100\,\mu\mathrm{s}$, LONG used $100,200,300,400,500\,\mu\mathrm{s}$, and MIXED used $20,60,100,300,500\,\mu\mathrm{s}$. Each sequence had two acquisition periods, A/B, at two targets per size. Here MIXED denotes annealing durations, not heterogeneous chains. The $L_{\mathrm b}=36$ study included SHORT and LONG only. Four gauges and two repeats of 500 reads gave eight QMIs per condition. The complete comparison contains 160 conditions, 1280 QMIs, and 640000 reads.

The $L_{\mathrm b}=6,19$ maps remained valid across the working-graph change between SHORT and LONG/MIXED. The repaired $L_{\mathrm b}=36$ map was validated on the updated graph, but inherited scales from a different calibration map. Structural validity does not revalidate temperature matching. These single-map time-study representatives are distinct from the later parallel banks and establish no serial--parallel distribution equivalence. Within-period contrasts hold mapping and scale fixed. Cross-period contrasts additionally include acquisition variability and, where applicable, graph changes.

\begin{figure*}[t]
\centering
\includegraphics[width=\textwidth]{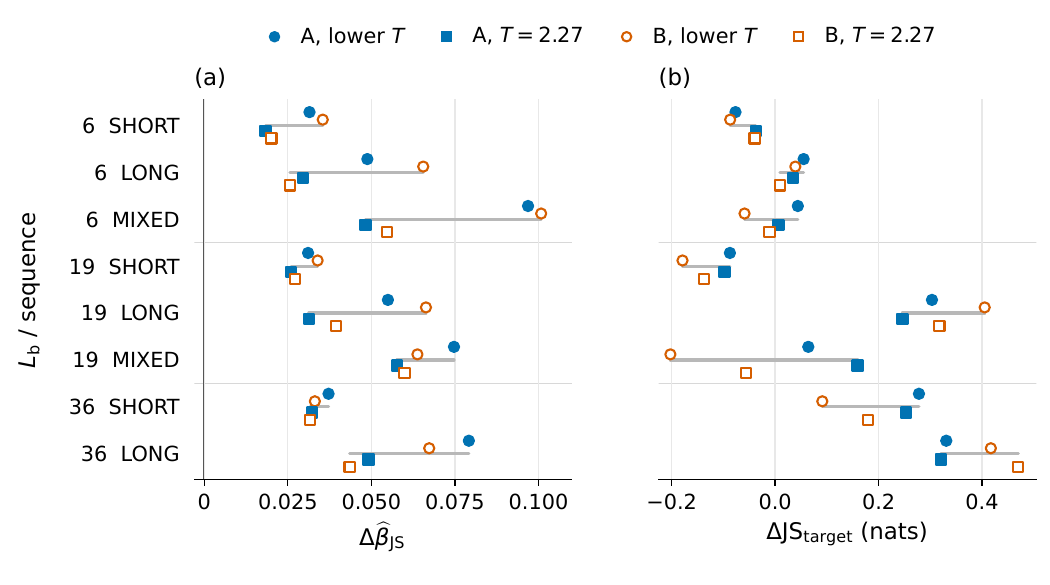}
\caption{Fixed-scale annealing endpoint changes for chain-free blocks: (a) fitted inverse temperature and (b) target energy-distribution JS divergence. Each point is a within-period longest-minus-shortest contrast: $100-20\,\mu\mathrm{s}$ (SHORT), $500-100\,\mu\mathrm{s}$ (LONG), or $500-20\,\mu\mathrm{s}$ (MIXED). Filled blue markers denote period A. Open orange markers denote period B. Circles indicate the lower target ($T=1.82,2.12,2.17$ for $L_{\mathrm b}=6,19,36$). Squares indicate $T=2.27$. Gray spans are observed ranges, not confidence intervals. All 32 inverse-temperature changes are positive. Negative target-JS changes denote improved agreement.}
\label{fig:new-anneal-summary}
\end{figure*}

Figure~\ref{fig:new-anneal-summary} reduces the results to their within-period endpoints. All 32 fitted inverse-temperature changes are positive, ranging from 0.01833 to 0.10081, and their reported pointwise QMI intervals exclude zero. Mean absolute magnetization also increases in every contrast. This is neither a simultaneous confidence statement nor evidence of monotonicity at every intermediate duration.

Larger fitted inverse temperature does not uniformly improve target agreement. For $L_{\mathrm b}=6,19$, target-energy JS divergence decreases in all eight SHORT contrasts but increases in all eight LONG contrasts. MIXED increases in all four A-period contrasts and decreases in all four B-period contrasts. For $L_{\mathrm b}=36$, all eight contrasts increase the target mismatch. Thus even the direction of improvement can depend on acquisition period, and longer annealing cannot be ranked by temperature shift alone.

The repaired-map $L_{\mathrm b}=36$ estimates agree to numerical precision between two qualified finite-OBC references, yet the target lies outside 38 of their 40 reported JS intervals. Reference agreement therefore does not imply successful calibration transfer. Magnetization and Binder residuals remain separate tests, and a fitted energy distribution does not determine the full spin law. Together, these controls motivate choosing an operating duration together with its calibrated scale, physical map, and independent distribution checks. Neither $100\,\mu\mathrm{s}$ nor the longest measured duration is a universal optimum.

\section{From block distributions to corrected sampling on larger periodic lattices}
\label{sec:sampling}

For full-lattice observables below, the densities are $e=H(s)/V$ and $m=M/V$, with $V=L^2$, rather than the block-normalized quantities used in the preceding comparisons.

We distinguish two uses of the measured blocks. The archived $L_{\mathrm b}=19$ benchmarks generate candidates from an evaluable learned distribution and apply exact MH/MTM correction. The later direct-QPU experiment generates candidates on the hardware and tests an approximate temperature-based ratio. Their training samples, boundary conditions, and acquisition histories remain separate.

\subsection{Normalized proposals and exact correction}

An autoregressive proposal assigns a normalized microstate probability,
\begin{equation}
 q_\theta(s_S\mid b)=\prod_{k=1}^{N_{\mathrm b}}q_\theta(s_k\mid s_{<k},b).
 \label{eq:ar}
\end{equation}
Here $k$ indexes the spin-generation order, $s_{<k}$ contains the preceding spins, $b$ is the fixed boundary input, and $\theta$ denotes the learned parameters. The boundary-conditioned GRU has 128 hidden components, two dense boundary-encoding layers, and raster-order inputs containing previous, left and above spins, boundary fields, and coordinates. Maximum-likelihood training uses adaptive moment estimation with decoupled weight decay (AdamW) for eight epochs and selects the lowest validation NLL checkpoint. This full-configuration autoregressive likelihood differs from the sitewise conditional pseudolikelihood used for thermometry. Contexts, rather than individual reads, are split into training/validation/test sets: 40/5/5 in the original panel and 21/3/6 in the refined mixed panel, with 1000 reads per context. Finite logits give full support. Stable log probabilities avoid numerical underflow. The temperature scan below uses a separate boundary-independent family $q_{\theta,T}(s_S)$.

For $y\sim q_\theta(\cdot\mid b)$ replacing $x$, independence MH accepts with~\cite{metropolis1953,hastings1970}
\begin{equation}
 \begin{aligned}
 A(x\to y)=\min\bigg[1,{}&e^{-\beta[H_S(y;b)-H_S(x;b)]}\\[-2pt]
 &\times\frac{q_\theta(x\mid b)}{q_\theta(y\mid b)}\bigg].
 \end{aligned}
 \label{eq:mh}
\end{equation}
Candidates must be generated from the same frozen model whose probabilities enter the ratio. With full support and the complete boundary energy, this preserves the specified target regardless of proposal error. A fitted hardware temperature can define a benchmark target before evaluation. It does not change that target during a chain. Substituting a fitted probability for an unknown raw-QPU probability is a different algorithm.

For MTM, $K$ denotes the number of candidates, distinct from the fitted pair coefficients $K_{ij}$. Draw $K$ independent forward candidates from $q_\theta(\cdot\mid b)$, define $w(z)=e^{-\beta H_S(z;b)}/q_\theta(z\mid b)$, and choose $y_J$ with probability proportional to $w(y_J)$. Independently draw $K-1$ reverse candidates $x_j^\star$ from the same proposal and include the old state $x$ in the reverse set. The acceptance is
\begin{equation}
 A_{\mathrm{MTM}}(x\to y)=\min\left[1,
 \frac{\sum_{j=1}^{K}w(y_j)}{w(x)+\sum_{j=1}^{K-1}w(x_j^\star)}\right].
 \label{eq:mtm}
\end{equation}
Sums are evaluated logarithmically. $K=1$ recovers independence MH~\cite{liu2000}. Increasing $K$ can improve acceptance while increasing proposal-evaluation work.

\subsection{Temperature and learning controls}
\label{sec:temperature-acceptance}
\label{sec:classical-learning-control}

\begin{figure*}[t]
\centering
\includegraphics[width=0.96\textwidth]{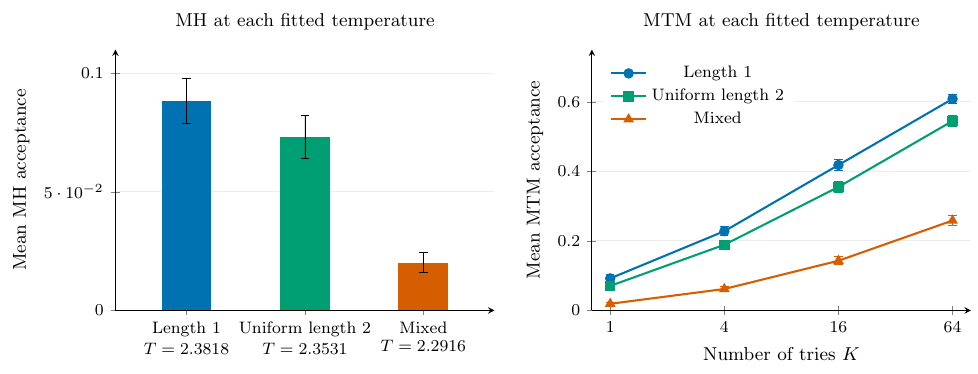}
\caption{MH and MTM acceptance on 2048 equilibrium blocks at each proposal's own fitted temperature. MTM varies the candidate count $K$. Error bars are 95\% intervals from grouped resampling of evaluation records, with models and fitted temperatures fixed. Unlike Table~\ref{tab:uniform-temperature}, the targets differ between proposals.}
\label{fig:acceptance}
\end{figure*}

Acceptance tests use 2048 equilibrium blocks from four Wolff trajectories, with uncertainty estimated by resampling consecutive groups within each trajectory. Models and fitted temperatures remain fixed, so these intervals exclude refitting uncertainty. At each model's own $T_{\mathrm{fit}}=1/\widehat\beta_{\mathrm{PL}}$, the chain-free and uniform-length-two proposals outperform the mixed proposal (Fig.~\ref{fig:acceptance}). This tests learned-proposal overlap after a temperature adjustment. Table~\ref{tab:uniform-temperature} instead holds the target temperature common to all proposals. It retains residual temperature mismatch and is a complementary control. Acquisition-period differences prevent attributing either ordering to chain length alone. Neither test establishes production-chain convergence.

\begin{table}[t]
\centering
\caption{Acceptance on common equilibrium contexts at $T=2.3531$, the uniform-length-two conditional fitted temperature. The mixed samples come from a separate acquisition.}
\label{tab:uniform-temperature}
\small
\begin{tabular}{lrrr}
\toprule
Embedding & Mean $\ell$ & MH & MTM ($K=64$)\\
\midrule
Chain-free & 1.000 & 0.0849 & 0.5749\\
Uniform-length-two & 2.000 & 0.0731 & 0.5447\\
Mixed-chain & 1.504 & 0.0369 & 0.3432\\
\bottomrule
\end{tabular}
\end{table}

Coarse scans around the fitted temperatures give slightly higher acceptance above the fitted value for C2 and mixed proposals, without resolving an optimum. Thermometry therefore defines a diagnostic reference, not necessarily the temperature of maximal acceptance.

Classical-trained controls isolate rejection that can occur without QPU training data. They use the same GRU at $T_c$ and 2.5, with disjoint training and evaluation boundaries. Four heat-bath chains per boundary pass the recorded energy and magnetization checks (split convergence diagnostic $\widehat R\leq1.05$ and effective sample size, ESS$\geq200$). These study-specific screens are less stringent than the $\widehat R<1.01$ recommendation of Ref.~\cite{vehtari2021}. They are not a universal convergence certificate. Primary MH/MTM acceptance is 0.1776/0.6798 at $T_c$ and 0.1823/0.7395 at 2.5. Increasing training resources improves the tested values (Table~\ref{tab:classical-learning}). These are learning controls, not an additive subtraction of hardware error, because the classical and QPU panels use different contexts.

\subsection{Temperature scan and critical relaxation}

\begin{figure*}[t]
\centering
\includegraphics[width=0.96\textwidth]{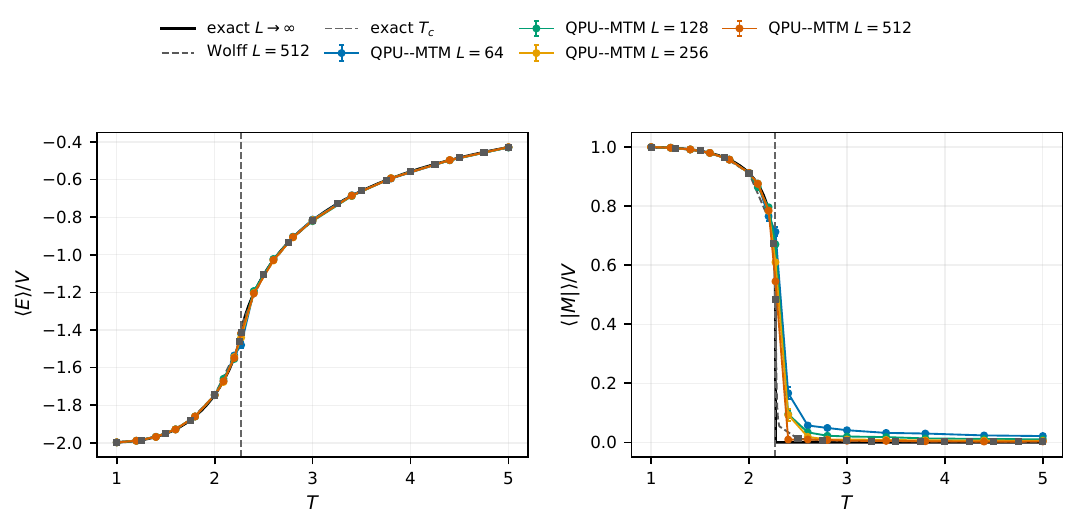}
\caption{Energy and absolute magnetization densities from temperature-matched, exactly corrected MTM for $L=64$--512. Heat-bath and Wolff updates prepare the states. Production contains no Wolff updates. The grey dashed curve shows the $L=512$ Wolff reference and black solid curves the thermodynamic limit. Error bars account for autocorrelation. Finite-volume absolute magnetization remains nonzero at $T_c$.}
\label{fig:temperature}
\end{figure*}

The prepared-ensemble scan uses 17 temperatures spanning $T=1$--5 and $L=64,128,256,512$. At each temperature, a boundary-independent GRU is trained on 8000 of 10000 zero-field $19\times19$ OBC reads, with 1000 each for validation and testing. Spin reversal augments training only. Current exterior spins enter every forward and reverse MTM energy. Thus the model generates proposals classically without sending changing boundaries to the QPU. Lattices receive 1000 checkerboard heat-bath sweeps and Wolff thermalization before MTM-only production. Figure~\ref{fig:temperature} compares the resulting observables with finite-volume Wolff and thermodynamic-limit references~\cite{onsager1944,yang1952}. At finite $L$, $\langle|m|\rangle$ need not vanish at $T_c$. This prepared scan tests observable agreement, not relaxation from an arbitrary initial state.

The scan uses $K=16$ and one trajectory per temperature and volume. For $L=64,128,256,512$, respectively, it records 1500, 1200, 700, and 350 block sweeps, with one measurement after each sweep and no additional MTM burn-in discard. Wolff preparation stops when its cumulative flipped-spin count reaches $50L^2$ or 5000 cluster updates, whichever occurs first. The latter limit can terminate preparation before 50 volume units, particularly at high temperature. The plotted standard errors use $s_O\sqrt{2\tau_{\mathrm{int},O}/n}$, where $s_O$ is the sample standard deviation and $n$ the number of records. The autocorrelation sum stops at the first nonpositive lag or when the lag exceeds six times the running integrated time. These single-trajectory error estimates do not measure uncertainty from initialization or independent repetitions.

Reusing the $T_c$-trained proposal at higher temperatures instead causes acceptance to collapse and energy estimates to retain initialization dependence. Exact correction cannot rescue a finite trajectory with almost no accepted moves. Temperature-matched models avoid that overlap failure in the tested scan. The heat-bath preparation is also important at high temperature, where small Wolff clusters alone may not erase the initial condition.

\begin{figure*}[t]
\centering
\includegraphics[width=0.96\textwidth]{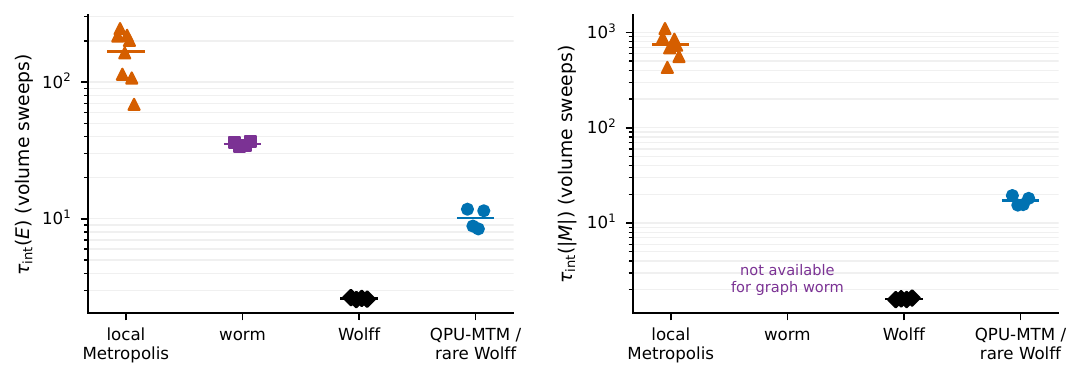}
\caption{Integrated autocorrelation at $L=128,T_c$ in site-work volume units. The hybrid combines MTM blocks and Wolff updates. Points represent independent runs. Horizontal lines are means. The logarithmic scale compares update work, not elapsed time. Worm records have no configuration-level absolute magnetization.}
\label{fig:volume-autocorrelation}
\end{figure*}

\begin{figure*}[t]
\centering
\includegraphics[width=0.96\textwidth]{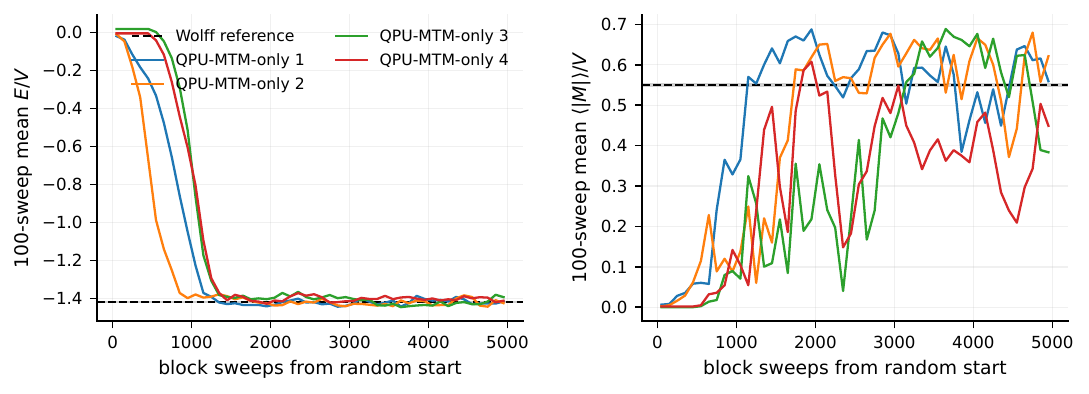}
\caption{Block-only MTM from four random starts at $L=128,T_c$, with $K=16$. Curves average 100 block sweeps. The band is the Wolff reference uncertainty. Energy relaxes faster than magnetization, which remains inconsistent across chains after 5000 sweeps.}
\label{fig:qpu-only}
\end{figure*}

Figure~\ref{fig:volume-autocorrelation} compares critical relaxation in volume-work units. At $L=128,T_c$, four hybrid chains use $K=16$ and one Wolff update per ten block sweeps, recording 5000 sweeps and discarding the first 1000. A block sweep attempts 36 nonoverlapping $19\times19$ replacements at random spatial offsets. Energy and magnetization agree across chains and with the Wolff reference within combined errors. In contrast, four random-start block-only chains remain nonstationary in magnetization after 5000 sweeps, although energy approaches the reference (Fig.~\ref{fig:qpu-only}). Two equilibrium-start chains of 1500 sweeps also retain magnetization offsets larger than their individual standard errors. For this critical-relaxation comparison, autocorrelation errors are reported only for traces passing stationarity checks~\cite{vehtari2021}.

For comparison with Metropolis, worm~\cite{prokofev2001}, and pure Wolff updates, we express autocorrelation in site-work units. A Metropolis sweep attempts all $V=L^2$ sites. Worm and Wolff times are multiplied by the mean attempted worm length or flipped-cluster size divided by $V$. Hybrid work is $(N_{\mathrm b}n_{\mathrm{block}}+n_{\mathrm{flip}})/V$, where $n_{\mathrm{block}}$ counts attempted block updates and $n_{\mathrm{flip}}$ counts spins flipped by Wolff updates over the same recording interval. Every worm or cluster is recorded separately. The Metropolis baseline has eight chains with 20000 thermalization and 100000 measurement sweeps. Worm and Wolff each have four chains with 250000 production records, 20000 thermalization records, and a further 5000-record analysis discard. Worm configurations do not supply a spinwise $|M|$ observable.

The hybrid decorrelates faster than local Metropolis, but pure Wolff is fastest (Fig.~\ref{fig:volume-autocorrelation}). Wolff contributes 3.95\% of the hybrid site work, so the improvement cannot be assigned to QPU-informed blocks alone. Site work is not wall time, neural-inference cost, or QPU access time.

\subsection{Direct QPU block updates near criticality}
\label{sec:new-fbc-direct}

The direct-QPU experiment instead generates each candidate on the hardware for its current fixed boundary. It uses $L=128$, $L_{\mathrm b}=6,19,36$, $t_a=20,100\,\mu\mathrm{s}$, and
\begin{equation}
 T\in\{2.17,2.22,T_c,2.32,2.37\},
 \qquad T_c=\frac{2}{\ln(1+\sqrt2)}.
 \label{eq:new-fbc-temperature-grid}
\end{equation}
Fresh-OBC scale selections fix the programmed $\alpha$, while mapping-specific conditional-PL fits fix the proposal-temperature estimate $\widehat\beta_{q,k}$. These are separate calibration inputs and neither is updated during production. Each archived-OBC initialization undergoes checkerboard block updates, retaining all 20 successive configurations, including repeats, without burn-in removal or thinning. Two acceptance rules yield 60 one-chain conditions, 1200 configurations, and 219600 block attempts. No neural proposal is used. One retained configuration is a full-lattice record after a sequence of block attempts, not one QPU read or one accepted block. The attempts per record depend on block size (Table~\ref{tab:new-resource-occupancy}).

Writing $\Delta H_S=H_S(y;b)-H_S(x;b)$, the energy-only rule omits the unknown hardware ratio. The \texttt{on} rule substitutes a fitted Gibbs ratio:
\begin{align}
 \log A_{\mathrm{off}}&=\min(0,-\beta\Delta H_S),\nonumber\\
 \log A_{\mathrm{on}}&=\min(0,-\Delta\beta_{\mathrm{acc},k}\Delta H_S),\nonumber\\
 \Delta\beta_{\mathrm{acc},k}&=\beta-\widehat\beta_{q,k}.
 \label{eq:new-fbc-acceptance}
\end{align}
Here $k$ labels the physical block mapping used for the attempt, and $\widehat\beta_{q,k}$ is its frozen conditional-PL estimate at the chosen scale and annealing time. The mismatch is target minus proposal inverse temperature. Mapping-specific estimates are monitored per attempt, not refitted during the chain. Their fold spread is not a confidence interval or a hard exclusion threshold.

For one fixed context, the \texttt{on} rule would be exact if the actual proposal were the fitted Gibbs law. A scalar fit does not establish this. The parallel schedule excludes overlapping or nearest-neighbour-connected blocks and checks energy additivity, but full-target preservation also requires the correct joint proposal law. Disjoint physical resources do not establish statistical independence. For an independent raw proposal $q$, \texttt{off\_energy} instead preserves a conditional law proportional to $\pi_\beta q$. The fitted ratio removes only the assumed Gibbs factor, leaving any non-Gibbs residual (Appendix~\ref{app:new-fbc-diagnostics}).

\begin{figure*}[t]
\centering
\includegraphics[width=\textwidth]{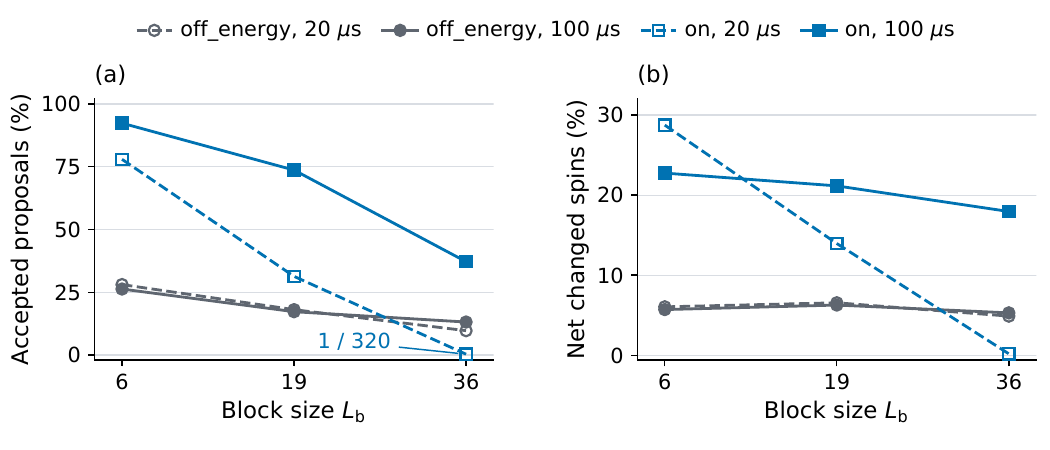}
\caption{Direct-QPU FBC diagnostics at $L=128,T_c$: block acceptance and the mean fraction of full-lattice spins changed per recorded update. Gray circles denote \texttt{off\_energy}. Blue squares denote \texttt{on}. Open markers with dashed lines denote $20\,\mu\mathrm{s}$, and filled markers with solid lines denote $100\,\mu\mathrm{s}$. Each condition contains one chain of 20 retained configurations. The $L_{\mathrm b}=36$, $20\,\mu\mathrm{s}$ \texttt{on} trajectory accepts only one of 320 proposals. The plotted quantities measure observed transitions, not equilibrium efficiency. Detailed moments and counts are in Table~\ref{tab:new-fbc-critical}.}
\label{fig:new-fbc-summary}
\end{figure*}

All recorded acceptance decisions reproduce their energy term, correction, and random variate. On the same 109800 stored \texttt{on} attempts, removing only the correction would reduce acceptance from 76.64\% to 46.91\%. This isolates a decision-level effect. It does not reconstruct an energy-only trajectory. The aggregate is dominated by size-six proposals, so Fig.~\ref{fig:new-fbc-summary} emphasizes the block-size dependence and actual spin motion at $T_c$.

The most restrictive case is $L_{\mathrm b}=36$, $20\,\mu\mathrm{s}$ with the \texttt{on} rule: one of 320 proposals is accepted and 19 recorded updates leave the state unchanged. Here $\widehat\beta_q=0.41752024$ and $\Delta\beta_{\mathrm{acc}}=0.02316655$. All proposed energy increments are positive, with median 368 and corresponding acceptance about $2\times10^{-4}$. Acceptance depends on the signed product in Eq.~\eqref{eq:new-fbc-acceptance}, not on mismatch magnitude alone. Conversely, the size-six \texttt{on} chain has higher acceptance at $100\,\mu\mathrm{s}$ but less net spin change than at $20\,\mu\mathrm{s}$.

Observable traces also remain transient. All energy-only trajectories lower their mean energy between the first and last five records. Only two of 570 consecutive-record transitions increase it. The nearly frozen large-block case nevertheless gives $U_4=0.66469$, close to $2/3$, from only two visited states. Such moments, and strongly window-dependent fluctuation estimates, do not establish equilibrium. The 20-record results diagnose acceptance and mobility. They are not an equilibrium Binder comparison or a reliable critical autocorrelation estimate. Those comparisons require longer independent trajectories and a matched finite-$L$ reference.

\section{Discussion}
\label{sec:discussion}

\subsection{What representation and thermometry establish}

The present comparisons connect three established lines of work: representation-dependent sampling errors~\cite{marshall2020}, operational thermometry~\cite{benedetti2016,raymond2016}, and corrected neural proposals~\cite{scriva2023,wu2021}. Their joint experimental test is consequential because success at one level need not transfer to the next. Chain-free embeddings remove auxiliary-chain errors, but not distributional bias. Mean-energy calibration leaves differences in energy widths and magnetization, while uniform-length-two gives smaller magnetization deficits than the tested mixed embeddings. The conditional analysis in Sec.~\ref{sec:chain-range} finds predictive gains from longer-range terms in mixed embeddings after fitting individual nearest-neighbour couplings and fields, with no resolved benefit in the tested C1 and C2 controls. These effective dependencies are not additional programmed interactions or a microscopic noise diagnosis~\cite{vuffray2022}. Placement and acquisition period remain confounded with chain class, so the results support prioritizing C1 and short uniform chains without assigning a universal ranking.

Within C1, chip occupancy measures acquisition capacity rather than independent information. Small blocks have more boundary sites relative to their area. Large blocks propose more extensive rearrangements but can be almost immobile. With exact conditional or valid Hastings updates, every block size preserves the same full-lattice target. Isolated-OBC finite-size effects therefore do not automatically become a bias of the final periodic-boundary-condition (PBC) ensemble. The direct-QPU study instead combines boundary-response errors, approximate correction, and limited mixing.

Thermometry separates target displacement from remaining distributional discrepancy. All 32 endpoint changes in JS-fitted inverse temperature are positive, without uniform improvement in target agreement. Period variability and the graph/mapping changes limit cross-period causal attribution. Agreement between two qualified $L_{\mathrm b}=36$ reference analyses establishes robustness to those revisions, not an exact reference or successful calibration transfer. The uncertainty and acquisition controls for these comparisons are specified in Sec.~\ref{sec:thermometry-criteria} and Appendix~\ref{app:obc-evidence}.

\subsection{Correct target, acceptance, and convergence}

A macrostate histogram $P(E,M_S)$ cannot replace a microstate proposal probability. Generating from the same frozen normalized model whose probabilities enter MH/MTM makes exact correction possible~\cite{mcnaughton2020,wu2021,scriva2023}. Its quality then controls acceptance and mixing. The learned-proposal benchmarks show both prepared-ensemble agreement and slow critical magnetization relaxation. Hybrid results belong to the combined block-plus-Wolff kernel, not to QPU-informed blocks alone. Stein weighting offers another way to use annealer samples without an explicit source density~\cite{shibukawa2024}. It corrects sample weights and is distinct from the exactly invariant transitions considered here.

For raw-QPU proposals, removing the fitted-temperature term does not generally repair a temperature error, and adding it does not remove an unknown non-Gibbs residual. Same-attempt log-ratio diagnostics separate this arithmetic effect from differences between evolved trajectories. Acceptance must also be distinguished from accepted spin displacement. Neither a large accepted fraction nor a finite autocorrelation estimate from 20 records certifies equilibration.

Performance comparisons must respect the two cost structures: learned candidates are generated classically after training, whereas direct FBC consumes QPU acquisitions during the trajectory. QPU access, training, inference, queueing, and optimized classical baselines must be included. Occupancy, nominal anneal exposure, and site-work autocorrelation cannot replace missing access-time records. The present results establish controlled ways to use and assess annealer samples, rather than a computational advantage over optimized classical sampling.

\subsection{Scope beyond the square-lattice Ising benchmark}

The hardware audit, fixed-placement controls, comparison of operational temperatures, independent calibration checks, and microstate Hastings correction apply more broadly than this particular ferromagnet. Classical competitors must change with the model: dilution can impair Wolff updates while leaving Swendsen--Wang effective~\cite{kole2022}, and frustrated models admit parallel-tempering and isoenergetic-cluster controls~\cite{zhu2015}. For another theory or Gibbs target, its Hamiltonian, state-space support, boundary interactions, and reference calculations must be specified anew. The Ising form used here avoids auxiliary variables and penalties that some application mappings require.

In particular, a reduction that preserves ground-state optimization need not preserve finite-temperature probabilities after auxiliary variables are marginalized. Extending the method therefore requires checking the induced distribution, not only the encoded energy minimum. Exactly corrected learned proposals require nonnegative target weights, evaluable ratios, and the same normalized proposal for generation and probability evaluation. A sign or complex-weight problem would need an additional treatment. The transferable result is this validation and correction procedure. Accuracy, mixing, and speed must be established for each model~\cite{nelson2022,pelofske2026,sathe2026,teza2026}.

\subsection{Practical implications for sampling}

The results suggest a sequence of tests. Start with one physical qubit per logical spin when feasible, and evaluate short uniform chains otherwise. Calibrate the input scale, then test magnetization, energy fluctuations, and conditional correlations on independent data. For learned proposals, use the same frozen normalized model for generation and probability evaluation, and assess full-lattice relaxation as well as acceptance. For direct-QPU proposals, validate the approximation to the proposal ratio separately. This sequence connects embedding selection to sampling without treating chain integrity, temperature matching, or acceptance alone as sufficient.

\section{Conclusions}
\label{sec:conclusions}

We connect validated hardware representation, Gibbs-distribution tests, and full-lattice sampling in a two-dimensional Ising benchmark. Fixed one-to-one mappings reach $36\times36$ spins without chains. The $19\times19$ comparison with uniform-length-two and mixed-chain embeddings shows that removing chain breaks and matching mean energy do not ensure agreement of magnetization or sampled correlations. Mixed embeddings retain a held-out predictive benefit from extended-range conditional terms, with no resolved benefit in the tested C1/C2 controls. These results distinguish representation effects, temperature displacement, and remaining distributional errors.

Parallel searches provide verified banks of 97 size-six and seven size-nineteen blocks without certifying global maximality. The completed fixed-scale $20$--$500\,\mu\mathrm{s}$ scans show systematic endpoint temperature shifts, but target agreement depends on duration, size, and acquisition period. Direct-QPU fixed-boundary updates retain 1200 configurations across five temperatures, three block sizes, two durations, and two acceptance rules. Their acceptance and spin-displacement diagnostics expose sharply different dynamics, including an almost frozen large-block trajectory. These short records do not establish equilibrium critical observables or a preferred block size.

Normalized GRU proposals provide a separate route to exact MH/MTM updates on larger periodic lattices, making imperfect training samples usable without inheriting their distribution as the target. Correct invariant measure does not guarantee rapid equilibration: critical magnetization remains slow under block-only updates, whereas hybrid agreement relies on collective Wolff moves. Extension to other theories transfers the hardware controls and validation procedure, with each encoded Gibbs distribution and its dynamics checked anew. No quantum advantage is claimed.

\subsection*{Acknowledgements}
The authors acknowledge the J\"ulich Supercomputing Centre and access through the J\"ulich UNified Infrastructure for Quantum computing (JUNIQ) to quantum annealing resources from D-Wave. This research was supported by 'Creation of the quantum information science R\&D ecosystem(based on human resources)' through the National Research Foundation of Korea(NRF) funded by the Korean government (RS-2023-NR057243).
This research was supported by the National Research Foundation (NRF) funded by the Korean government (MSIT) (No. RS-2025-02221606, RS-2022-NR070836).

\appendix

\section{Embedding and parallel resources}
\label{app:chainfree-search}

The search uses the active hardware graph and validates every one-qubit image and required grid bond. It is non-induced: extra hardware edges are allowed but not programmed. For a target window $W$, $|W|\ge L_{\mathrm b}^2$ and $|C[W]|\ge2L_{\mathrm b}(L_{\mathrm b}-1)$ are necessary, not sufficient, capacity conditions. Small-grid backtracking excludes used qubits and intersects the neighbourhoods of assigned logical neighbours,
\begin{equation}
 D_v(A)=(D_v^{(0)}\setminus\varphi(A))
 \cap\!\bigcap_{u\in N(v)\cap A}N_{\mathrm{hw}}(\varphi(u)),
 \label{eq:search-domain}
\end{equation}
Here $A$ is the set of already assigned logical vertices, $D_v(A)$ is the remaining candidate set for vertex $v$, and $D_v^{(0)}$ contains degree-compatible qubits. The sets $N(v)$ and $N_{\mathrm{hw}}(q)$ are logical and hardware neighbours, respectively. In the capacity condition, $C[W]$ denotes hardware couplers with both endpoints in $W$. Seeded tie-breaking and forward checking explore multiple placements. Glasgow portfolios vary ordering, windows, labels and optional cross/strip anchors. A partial anchor is accepted only after full-grid completion and independent validation~\cite{mccreesh2020glasgow}. Regional anchors remain constraints even when their windows expand, unlike unrestricted whole-graph attempts.

All 324 contiguous $19\times19$ crops of one $36\times36$ parent share four central sites and cannot form a disjoint bank. Parallel packing therefore combines separately placed candidates, checking qubit overlap and the configured separation. A validated mapping proves existence on that graph. Bounded search over a finite candidate pool does not establish global maximality. Fixed maps are reused throughout their comparisons. Geometry describes the graph drawing, not a measured quality ranking, and resource disjointness does not establish statistical independence of parallel responses.

\subsection{Occupancy and timing}
\label{app:new-resource-accounting}

The 105 maps in the $97+7+1$ banks pass graph, bond, injectivity and within-bank disjointness checks on the 4560-qubit active graph. Table~\ref{tab:new-resource-occupancy} distinguishes bank capacity from actual QMI occupancy. Size six uses 800 QMIs of 80 blocks and 1600 of 81. Size nineteen uses 2800 of six and 400 of seven. Size thirty-six uses 6400 single-block QMIs. The 12000 QMIs and 219600 proposals exclude OBC initialization and calibration.

\begin{table*}[t]
\centering
\caption{Parallel resources for the direct-QPU $L=128$ trajectories. Occupancy is relative to 4560 active qubits. ``Bank'' uses the entire selected bank. Mean QMI occupancy uses the blocks assigned to each request. Proposed-site work is $n_{\mathrm{prop}}L_{\mathrm b}^2/V$, $V=128^2$, including rejected attempts. The rows are alternative acquisitions, not one mixed-size packing.}
\label{tab:new-resource-occupancy}
\small
\setlength{\tabcolsep}{4pt}
\begin{tabular}{rrrrrrrrr}
\toprule
$L_{\mathrm b}$ & Bank maps & Bank qubits & Bank (\%) & Blocks/QMI & Mean QMI (\%) & Proposals/config. & QMIs/config. & Site work/$V$\\
\midrule
6 & 97 & 3492 & 76.58 & 80--81 & 63.68 & 484 & 6 & 1.06348\\
19 & 7 & 2527 & 55.42 & 6--7 & 48.49 & 49 & 8 & 1.07965\\
36 & 1 & 1296 & 28.42 & 1 & 28.42 & 16 & 16 & 1.26563\\
\bottomrule
\end{tabular}
\end{table*}

Proposed-site work can exceed one volume through repeated visits across update phases. It is distinct from accepted blocks, net Hamming change and independent information. The earlier GRU benchmark instead attempts 36 size-nineteen blocks per sweep, or $36\times361/128^2=0.79321$ volumes. For the fixed-scale annealing series, timing records cover only 640 LONG/MIXED QMIs for sizes six and nineteen, summing to 142.580532~s. The other 640 QMIs lack access-time records. Across all 1280 QMIs, read count times annealing duration totals 117.76~s, excluding programming/readout overhead. Neither this exposure nor the direct-FBC request count establishes complete QPU access time.

\section{Conditional models and classical references}
\label{app:conditional}
\subsection{Conditional fits and controls}
\label{app:range-fit}

For the C1, C2 and mixed-chain comparison in Sec.~\ref{sec:chain-range}, conditional log odds are $2(u_i+\sum_jK_{ij}s_j+\sum_aG_{ia}\sigma_a)$. Site fields are fitted individually, $K_{ij}=K_{ji}$, and exterior spins $\sigma_a$ enter only through programmed boundary bonds. $\sigma_a$ is distinct from the summed boundary field $b_i$. Starting from nearest neighbours, extensions to internal Manhattan distances two and three refit all coefficients. The limited-memory Broyden--Fletcher--Goldfarb--Shanno algorithm with bound constraints (L-BFGS-B) minimizes mean conditional NLL per spin plus $10^{-4}\|\vartheta\|_2^2/N_{\mathrm b}$, with $\vartheta$ containing all fitted conditional-model coefficients. These dimensionless coefficients absorb the effective inverse temperature and are distinct from the programmed physical couplings and biases.

Five-fold validation holds out complete boundary contexts, keeping all samples from one boundary together. Models fit four groups and predict the fifth, with identical splits across embeddings and the classical control. Predictive gains on unseen boundaries receive equal boundary weights. Their 95\% bootstrap intervals hold fitted models fixed, excluding refitting uncertainty and multiple-comparison correction. Earlier controls extend symmetric nearest-neighbour models by diagonal and axial distance-two terms. Checks vary predictors, splits, optimizer seeds and ridge coefficients $0,10^{-6},10^{-4}$. The independent initial C1/C2 comparison likewise finds no held-out benefit from the additional terms. These controls remain distinct from the mixed-chain result.

\paragraph{Chain-strength and time control.}
\label{app:intervention}
The intervention compares C1 and mixed placements at 5, 20 and $100\,\mu\mathrm{s}$, with mixed-chain strengths 0.5, 0.75 and 1. C1 has no internal chain coupling. Each of 60 conditions has a separate three-point scale calibration and 750 new reads on 24 common boundaries: 1080000 reads, with condition order randomized within boundaries. Scalar PL temperatures and QPU magnetization exclude broken-chain reads. Classical references simulate the original nearest-neighbour Hamiltonian at each condition's fitted temperature and the same boundaries. Four heat-bath chains per boundary start from positive, negative and two random states, discard 10000 sweeps and retain 4000 records each at 20-sweep intervals. Energy and signed/absolute magnetization satisfy split $\widehat R<1.05$ and ESS$\ge200$ on every boundary.

Each placement is compared with its own reference before equal weighting over placements and boundaries. The 95\% intervals use 20000 resamples of placements and shared boundaries, holding fitted temperatures and classical means fixed. Longer anneals reduce broken reads without comparably removing the magnetization deficit, which remains for C1. Recalibration at each condition distinguishes this intervention from the later fixed-scale time scans.

\subsection{Reference construction and support}
\label{app:obc-reference}

Small-block joint DOS checks include $\sum_Eg(E)=2^{N_{\mathrm b}}$, two ground states and the bipartite symmetry $g(E)=g(-E)$. Magnetization requires joint $g(E,M_S)$, not energy DOS alone. For larger blocks, construction and independent-check trajectories keep distinct roles: the reference-reuse revision assigns Wolff data to A and parallel-tempering checks to B. An independent checking sample is not merged into construction after inspecting disagreement.

For canonical histograms $H_a$, counts $n_a$ and statistical inefficiencies $\kappa_a$, WHAM uses~\cite{ferrenberg1989,kumar1992}
\begin{align}
 \widehat g(E)&=\frac{\sum_aH_a(E)/\kappa_a}
 {\sum_a(n_a/\kappa_a)e^{f_a-\beta_aE}},\label{eq:obc-wham}\\
 e^{-f_a}&=\sum_E\widehat g(E)e^{-\beta_aE}.\nonumber
\end{align}
Here $a$ labels a reference temperature, $H_a(E)$ counts observations at energy $E$, $n_a=\sum_EH_a(E)$, and $\beta_a$ is the reference inverse temperature. The fitted normalization is $f_a=-\log Z(\beta_a)$ up to the common DOS normalization. Numerical convergence, histogram connectivity, support, mixing and independent-observable agreement are separate checks. Correlated tempering temperatures are not independent validation runs. Reference-dependent fits are withheld when observed energies lack valid support. Raw QPU observations and empirical moments are retained. Covering observed levels alone does not validate unsampled tails.

Construction and checking data are compared using energy, absolute magnetization, Binder cumulant, distributional discrepancies, effective sample counts and mixing diagnostics over the validated temperature interval. Undefined diagnostics remain unavailable. In particular, the screening statistic $|e_{\mathrm b,\mathrm{WHAM},A}-\widehat e_{\mathrm b,B}|/\sqrt{\mathrm{SE}_A(e_{\mathrm b})^2+\mathrm{SE}_B(e_{\mathrm b})^2}$ reuses direct-MC errors. It is not a fully propagated WHAM uncertainty or sufficient reference validation.

\subsection{Estimator identities and uncertainty}
\label{app:obc-uncertainty}

For common support, weights and DOS,
\begin{align}
 D_{\mathrm{KL}}(\widehat P\Vert p_\beta)
 &=C_0+\beta\langle E\rangle_{\widehat P}+\log Z(\beta),\nonumber\\
 \partial_\beta D_{\mathrm{KL}}&=\langle E\rangle_{\widehat P}-\langle E\rangle_{p_\beta},\nonumber\\
 \partial_\beta^2D_{\mathrm{KL}}&=\operatorname{Var}_{p_\beta}(E)\ge0.\nonumber
\end{align}
Here $C_0$ is independent of $\beta$. Thus an interior forward-KL minimum equals the mean-energy match, not an independent thermometer. Reverse KL and JS need not share it. Separate energy-table interpolation can add numerical differences.

For $R_O=\widehat O-O_{\mathrm{ref}}(\widehat\beta)$, shared-QMI resampling retains the covariance in
\begin{align}
 \operatorname{Var}(R_O)\simeq{}&\operatorname{Var}(\widehat O)
 +[O'_{\mathrm{ref}}]^2\operatorname{Var}(\widehat\beta)\nonumber\\
 &-2O'_{\mathrm{ref}}\operatorname{Cov}(\widehat O,\widehat\beta).
 \label{eq:obc-shape-covariance}
\end{align}
The derivative $O'_{\mathrm{ref}}=\partial O_{\mathrm{ref}}/\partial\beta$ is evaluated at $\widehat\beta$. Parallel blocks remain grouped by physical QMI. Four gauges with two repeats yield only $3^4=81$ bootstrap multiplicity patterns. More draws do not add acquisitions. Intervals condition on the classical reference. Cross-fitting two repeat sets supplies held-out diagnostics but not independent repetitions or a propagated classical-reference error. Pooling periods also adds between-period variation,
\begin{equation}
 \operatorname{Var}_{\mathrm{pool}}(E)
 =\sum_rw_r\operatorname{Var}_r(E)+\sum_rw_r(\mu_r-\overline\mu)^2,
 \label{eq:obc-mixture-variance}
\end{equation}
where $r$ labels acquisition periods, $w_r$ is the fraction of observations from period $r$, $\mu_r$ is its mean energy, and $\overline\mu=\sum_rw_r\mu_r$, with $\sum_rw_r=1$. This identity does not establish a literal hardware-temperature fluctuation.

\paragraph{Gauge response.}
\label{app:obc-gauge}
Held-out site residuals $\langle s_i-\tanh(\widehat\beta_{-r}\eta_i)\rangle_r$ and symmetrized bond residuals are mapped to physical resources. Here $r$ labels an evaluation repeat, and $\widehat\beta_{-r}$ is fitted without that repeat. Gauge/repeat consistency diagnoses response anomalies. It does not identify defective hardware. Sector imbalance, sampling and collective distortions can contribute. A site without gauge-sign variation is unidentifiable, not unbiased. Balanced gauge/repeat sums of squares describe scatter about the measured mean. They do not assign causal noise fractions or detect a common bias.

\section{Calibration design and reproducibility}
\label{app:obc-evidence}

Independent size-six checks comprise four maps and three targets at $20\,\mu\mathrm{s}$. Eleven of twelve target inverse temperatures lie outside the reported JS intervals. These calibration-transfer failures remain in the analysis but do not estimate a universal failure probability. Mapping letters in that comparison denote placements, whereas A/B below denote acquisition periods.

Table~\ref{tab:obc-transfer} retains the selected size-nineteen and size-thirty-six independent checks. An interleaved low/candidate/high recheck brought the size-nineteen candidate response closer to its earlier calibration range. Those changing-scale, non-simultaneous acquisitions neither isolate geometry nor constitute a fixed-scale drift trace.

\begin{table*}[t]
\centering
\caption{Independent $20\,\mu\mathrm{s}$ calibration checks after JS-based interpolation. A/B identify maps. Intervals are 95\% intervals conditional on the supplied reference. These selected conditions illustrate transfer variability, rather than estimating a population failure rate.}
\label{tab:obc-transfer}
\small
\begin{tabular}{llrrrrl}
\toprule
$L_{\mathrm b}$ & Map & $T$ & $\alpha$ & $\beta$ & $\widehat\beta_{\mathrm{JS}}$ & Reported 95\% interval\\
\midrule
19 & A & 3.00 & 0.0370170136 & 0.333333 & 0.280666 & [0.278128, 0.283092]\\
19 & B & 3.00 & 0.0381716788 & 0.333333 & 0.285410 & [0.282402, 0.288391]\\
36 & A & 2.17 & 0.0572845239 & 0.460829 & 0.470282 & [0.468872, 0.471635]\\
36 & A & 2.27 & 0.0535771725 & 0.440529 & 0.440236 & [0.438463, 0.441789]\\
\bottomrule
\end{tabular}
\end{table*}

\paragraph{Scope of the earlier temperature scan.}
\label{app:obc-partial}
\label{sec:obc-coverage}
The earlier zero-field 30-target scan spans $T=1$--5 and includes 2.27 rather than exact $T_c$. It is distinct from the 17-temperature PBC scan and later five-temperature FBC study. Its complete L6 calibration conditions contain four gauges and two repeats, but the multi-size scan and independent calibration validation were incomplete. The observed energy-based temperature, magnetization and Binder discrepancies are target-reference comparisons, not held-out temperature-adjusted residuals or final PBC validation. Partial histograms weight observed gauges equally and their QMIs equally within gauge. A complete condition has the planned gauge/repeat set and can support its QMI interval even if other conditions are unfinished. An interpolation candidate additionally requires all planned probe points for that target to be complete and identifiable. Distinct acquisition periods remain separate. Full numerical data and acquisition identifiers are retained in the analysis archive.

\subsection{Fixed-scale time scans}
\label{new-anneal:appendix}

Table~\ref{new-anneal:protocol} gives the fixed scales. Table~\ref{new-anneal:endpoints} gives endpoint responses. The time scans use one fixed C1 map at each size. The size-six and size-nineteen maps differ from the corresponding FBC banks. The L36 map was repaired after a hardware-graph change, while its scale was inherited from a different map. Target equivalence therefore requires an independent check. Each condition has eight QMIs and 4000 reads. Six periods for L6/L19 and four for L36 yield 1280 distinct QMIs and 640000 reads. The two qualified L36 reference analyses reuse the same 320 QMIs, rather than adding independent observations. L6/L19 A-to-B midpoint separations are 152.591, 60.725 and 5.392 min for SHORT, LONG and MIXED, respectively. These characterize between-period comparisons, not QPU access time.

\begin{table}[t]
\centering
\caption{Frozen scales for the C1 time scans. S, L, M denote SHORT, LONG, MIXED, each acquired in A and B periods. A scale is fixed across all listed times and periods. Its availability does not certify target equivalence.}
\label{new-anneal:protocol}
\small
\begin{tabular}{crrl}
\toprule
$L_{\mathrm b}$ & $T$ & $\alpha$ & Sequences\\
\midrule
6 & 1.82 & 0.0560198243586226 & S, L, M\\
6 & 2.27 & 0.0432526839150392 & S, L, M\\
19 & 2.12 & 0.0494969380628002 & S, L, M\\
19 & 2.27 & 0.0459675653379849 & S, L, M\\
36 & 2.17 & 0.0572845238926354 & S, L\\
36 & 2.27 & 0.0535771724557724 & S, L\\
\bottomrule
\end{tabular}
\end{table}

\begin{table*}[t]
\centering
\caption{Endpoint responses: each row contains two target labels and two periods. Ranges span four point estimates, not confidence intervals. All 32 pointwise QMI intervals for $\Delta\widehat\beta_{\mathrm{JS}}$ exclude zero on the positive side. The last column counts reduced target-energy JS divergence. The MIXED improvements occur in B, not A.}
\label{new-anneal:endpoints}
\small
\begin{tabular}{clcrrc}
\toprule
$L_{\mathrm b}$ & Sequence & Endpoint ($\mu\mathrm{s}$) & $\Delta\widehat\beta_{\mathrm{JS}}$ range & $\Delta\langle|M_S|\rangle/N_{\mathrm b}$ range & Target-JS decrease\\
\midrule
6 & SHORT & $100-20$ & $[0.018333,0.035449]$ & $[0.027153,0.070653]$ & 4/4\\
6 & LONG & $500-100$ & $[0.025630,0.065522]$ & $[0.050917,0.108611]$ & 0/4\\
6 & MIXED & $500-20$ & $[0.048243,0.100812]$ & $[0.091903,0.216083]$ & 2/4\\
19 & SHORT & $100-20$ & $[0.025971,0.033918]$ & $[0.019978,0.054575]$ & 4/4\\
19 & LONG & $500-100$ & $[0.031271,0.066332]$ & $[0.143404,0.251917]$ & 0/4\\
19 & MIXED & $500-20$ & $[0.057641,0.074730]$ & $[0.222252,0.346443]$ & 2/4\\
36 & SHORT & $100-20$ & $[0.031668,0.037188]$ & $[0.135627,0.201690]$ & 0/4\\
36 & LONG & $500-100$ & $[0.043476,0.079171]$ & $[0.199975,0.220863]$ & 0/4\\
\bottomrule
\end{tabular}
\end{table*}

\section{Direct-QPU correction and trajectory statistics}
\label{app:new-fbc-diagnostics}
\label{app:obc-pbc-link}

For one fixed boundary and an independent actual proposal $q(s\mid b)$, the energy-only rule is MH for
\begin{equation}
 \widetilde\pi_{\mathrm{off}}(s\mid b)
 \propto q(s\mid b)e^{-\beta H_S(s;b)},
 \label{eq:new-fbc-off-law}
\end{equation}
since the proposal factors cancel in its Hastings ratio. If $q$ is Gibbs at $\beta_q$, this conditional law has inverse temperature $\beta+\beta_q$. For the fitted-temperature rule, write
\begin{equation}
 q(s\mid b)\propto e^{-\widehat\beta_qH_S(s;b)}r(s;b).
 \label{eq:new-fbc-proposal-residual}
\end{equation}
The corresponding auxiliary law is $\widetilde\pi_{\mathrm{on}}\propto q e^{-(\beta-\widehat\beta_q)H_S}\propto e^{-\beta H_S}r$: the fitted Gibbs factor is removed, but the residual is not. Bias in the fitted temperature is absorbed in $r$. These fixed-context identities do not imply a compatible global effective temperature for changing boundaries/mappings, nor the correct joint law for correlated simultaneous proposals. Exact learned-proposal MH instead draws from the same frozen normalized model used to evaluate the microstate ratio. A macrostate $P(E,M_S)$ or raw-QPU proxy is insufficient.

Table~\ref{tab:new-fbc-alpha} lists the programmed scales. Conditional-PL fits are separate, giving 1050 fixed mapping--condition values, $(97+7+1)\times5\times2$. Applied $\Delta\beta_{\mathrm{acc}}=\beta-\widehat\beta_q$ spans $-0.0341411$ to $0.0733528$. Fold spread is monitored without excluding proposals by a numerical threshold. An unavailable proposal-temperature estimate in the energy-only analysis does not mean zero mismatch.

\begin{table}[t]
\centering
\caption{Programmed FBC scales from finalized fresh-OBC selections, shown to ten decimals. Full precision remains in the acquisition plan. These amplitudes differ from the conditional-temperature fits used in acceptance and do not certify FBC temperature equivalence.}
\label{tab:new-fbc-alpha}
\scriptsize
\setlength{\tabcolsep}{3pt}
\begin{tabular}{rrrrr}
\toprule
$T$ & $t_a$ ($\mu$s) & $\alpha_6$ & $\alpha_{19}$ & $\alpha_{36}$\\
\midrule
2.17 & 20 & 0.0534068605 & 0.0556693162 & 0.0589606506\\
2.17 & 100 & 0.0509210973 & 0.0509716221 & 0.0534248656\\
2.22 & 20 & 0.0521163203 & 0.0542663095 & 0.0572003166\\
2.22 & 100 & 0.0497661718 & 0.0499797649 & 0.0524725167\\
$T_c$ & 20 & 0.0508901758 & 0.0528728552 & 0.0555443549\\
$T_c$ & 100 & 0.0486797233 & 0.0490467161 & 0.0515070577\\
2.32 & 20 & 0.0496754631 & 0.0514952921 & 0.0539072780\\
2.32 & 100 & 0.0476056648 & 0.0481243079 & 0.0503204978\\
2.37 & 20 & 0.0485310681 & 0.0501974739 & 0.0523649687\\
2.37 & 100 & 0.0465937818 & 0.0472013134 & 0.0492026260\\
\bottomrule
\end{tabular}
\end{table}

Table~\ref{tab:new-fbc-critical} retains all critical-temperature moments. Each condition is one 20-record trajectory. Repeated states after rejection remain in the sample. The moment formulas
\begin{align}
 C_V/V&=V\beta^2(\langle e^2\rangle-\langle e\rangle^2),\nonumber\\
 \chi/V&=V\beta(\langle m^2\rangle-\langle m\rangle^2),\nonumber\\
 \chi_{|m|}/V&=V\beta(\langle m^2\rangle-\langle|m|\rangle^2)
 \label{eq:new-fbc-fluctuation-moments}
\end{align}
also count drift in a nonstationary trace. Tail windows expose that sensitivity rather than replace the retained sample. At $T_c$, L36, $20\,\mu\mathrm{s}$, the \texttt{on} trace has eleven copies of one state and nine of another. $C_V/V=0.06778$ and $U_4=0.66469$ coexist with only one accepted proposal. Finite numerical autocorrelation outputs from these short traces do not establish equilibrium or between-chain convergence.

\begin{table*}[t]
\centering
\caption{Direct-QPU records at $L=128$, exact $T_c$. Each row contains 20 consecutive configurations without burn-in removal or thinning. $A$ is block acceptance. $d$ is mean net spin change, including initialization-to-first-record. Moments use all records and are not certified equilibrium estimates. Attempts per row are 9680, 980, 320 for $L_{\mathrm b}=6,19,36$. No equilibrium error bars are assigned.}
\label{tab:new-fbc-critical}
\small
\setlength{\tabcolsep}{7pt}
\begin{tabular}{rrlrrrrr}
\toprule
$t_a$ ($\mu$s) & $L_{\mathrm b}$ & Rule & $A$ (\%) & $d$ (\%) & $\langle e\rangle$ & $\langle|m|\rangle$ & $U_4$\\
\midrule
20 & 6 & \texttt{off\_energy} & 28.048 & 6.040 & -1.64113 & 0.17342 & 0.54966\\
20 & 6 & \texttt{on}          & 77.965 & 28.739 & -1.10621 & 0.17970 & 0.52856\\
20 & 19 & \texttt{off\_energy} & 18.061 & 6.547 & -1.47380 & 0.62893 & 0.61873\\
20 & 19 & \texttt{on}          & 31.327 & 13.950 & -1.16228 & 0.31896 & 0.63556\\
20 & 36 & \texttt{off\_energy} & 9.688 & 4.881 & -1.39279 & 0.22598 & 0.48918\\
20 & 36 & \texttt{on}          & 0.3125 & 0.174 & -1.32371 & 0.45771 & 0.66469\\
\midrule
100 & 6 & \texttt{off\_energy} & 26.281 & 5.698 & -1.64589 & 0.13121 & 0.59692\\
100 & 6 & \texttt{on}          & 92.366 & 22.745 & -1.38384 & 0.39213 & 0.56105\\
100 & 19 & \texttt{off\_energy} & 17.245 & 6.263 & -1.41146 & 0.50189 & 0.62385\\
100 & 19 & \texttt{on}          & 73.673 & 21.148 & -1.45038 & 0.63205 & 0.65293\\
100 & 36 & \texttt{off\_energy} & 13.125 & 5.287 & -1.50348 & 0.58365 & 0.64256\\
100 & 36 & \texttt{on}          & 37.188 & 17.948 & -1.28314 & 0.35472 & 0.58641\\
\bottomrule
\end{tabular}
\end{table*}

\section{Learning-control settings}
\label{app:learning-controls}

For the classical-trained control, four heat-bath chains per boundary start up, down, and from two random states, discard 10000 sweeps, and save 1000 states each at intervals of 100 sweeps. The selected 1000 states per boundary are balanced across chains. Twenty contexts train the GRU, five select the checkpoint, and ten are held out. Acceptance uses the final five with 100 separate reference states per context. The primary setting uses 128 hidden units, eight epochs, and three training seeds. The sensitivity settings in Table~\ref{tab:classical-learning} use one seed and were not selected by maximizing acceptance.

\begin{table*}[!t]
\centering
\caption{Classical-trained proposal acceptance on five held-out boundaries. Primary values average three seeds. Other settings use one. MTM uses $K=64$.}
\label{tab:classical-learning}
\small
\setlength{\tabcolsep}{4pt}
\begin{tabular}{llrr}
\toprule
$T$ & Training setting & MH & MTM\\
\midrule
$T_c$ & Primary & 0.1776 & 0.6798\\
 & 250 reads/context & 0.0945 & 0.5349\\
 & 32 epochs & 0.2023 & 0.7042\\
 & 256 hidden units & 0.1938 & 0.7215\\
\midrule
2.5 & Primary & 0.1823 & 0.7395\\
 & 250 reads/context & 0.1043 & 0.5993\\
 & 32 epochs & 0.2638 & 0.7602\\
 & 256 hidden units & 0.2300 & 0.7761\\
\bottomrule
\end{tabular}
\end{table*}

QPU-trained proposal acceptance is evaluated on 2048 equilibrium blocks and boundary contexts from four periodic-$128\times128$ Wolff chains, two ordered and two random. After 5000 cluster updates, each saves 512 configurations separated by 32 cluster updates. Group-bootstrap comparisons use common groups of 32, with 16 and 64 as sensitivity checks. Fitted temperatures and trained models are fixed during this resampling. The own-temperature tests used input scales 0.0622, 0.0910392, and 0.0622 for C1, C2, and mixed respectively. The mixed data were acquired in a different period.

\FloatBarrier

\end{document}